\documentclass[11pt]{article}
\pdfoutput=1

\usepackage[utf8]{inputenc}
\usepackage{times}

\usepackage{fullpage}

\usepackage[parfill]{parskip}

\usepackage{amsmath}
\usepackage{amssymb}
\usepackage{graphicx}
\usepackage{cite}
\usepackage{authblk}
\usepackage{xcolor}
\usepackage[unicode=true]{hyperref}

\title{Limits to causal inference with state-space reconstruction for infectious disease}

\author[1]{Sarah Cobey\thanks{cobey@uchicago.edu}}
\author[1]{Edward B. Baskerville\thanks{edbaskerville@uchicago.edu}}
\affil[1]{Ecology \& Evolution, University of Chicago, Chicago, IL, USA}
\date{}

\newcommand{\beginsupplement}{%
\setcounter{table}{0}
\renewcommand{\thetable}{S\arabic{table}}%
\setcounter{figure}{0}
\renewcommand{\thefigure}{S\arabic{figure}}%
}

\newcommand{\bx}{\mathbf{x}}
\newcommand{\hY}{\hat{Y}}

\begin{document}

\maketitle

\begin{abstract}

Infectious diseases are notorious for their complex dynamics, which make it difficult to fit models to test hypotheses.
Methods based on state-space reconstruction have been proposed to infer causal interactions in noisy, nonlinear dynamical systems.
These ``model-free'' methods are collectively known as convergent cross-mapping (CCM).
Although CCM has theoretical support, natural systems routinely violate its assumptions.
To identify the practical limits of causal inference under CCM, we simulated the dynamics of two pathogen strains with varying interaction strengths.
The original method of CCM is extremely sensitive to periodic fluctuations, inferring interactions between independent strains that oscillate with similar frequencies.
This sensitivity vanishes with alternative criteria for inferring causality.
However, CCM remains sensitive to high levels of process noise and changes to the deterministic attractor.
This sensitivity is problematic because it remains challenging to gauge noise and dynamical changes in natural systems, including the quality of reconstructed attractors that underlie cross-mapping.
We illustrate these challenges by analyzing time series of reportable childhood infections in New York City and Chicago during the pre-vaccine era.
We comment on the statistical and conceptual challenges that currently limit the use of state-space reconstruction in causal inference.

\end{abstract}

\section*{Background}

Identifying the forces driving change in natural systems is a major goal in ecology.
Because experiments are often impractical and come at the cost of generalizability, a common approach is to fit mechanistic models to observations.
Testing hypotheses through mechanistic models has a particularly strong tradition in infectious disease ecology \cite{KeelingRohani, AndersonMay, Kermack1927, Ross1910}.
Models that incorporate both rainfall and host immunity, for example, better explain patterns of malaria than models with only rainfall \cite{Laneri2010}; models with school terms fit the historic periodicity of measles in England and Wales \cite{Finkenstadt2000, Fine1982}.
The ability of fitted mechanistic models to predict observations outside the training data strongly suggests that biological insight can be gained. There is nonetheless a pervasive risk that predictive variables merely correlate with the true, hidden variables, or that the model's functional relationships create spurious resemblances to the true dynamics.
This structural uncertainty in the models themselves limits inference \cite{BurnhamAnderson, He2009, Yodzis1988, Wood1999, Grad2012}.

An alternative approach to inferring causality is to examine the time series of potentially interacting variables without invoking a model.
These methods face a similar challenge: they must distinguish correlated independent variables sharing a mutual driver from correlations arising from direct or indirect interactions.
Many of these methods, including Granger causality~\cite{Granger1969} and other related methods~\cite{Schumacher2015,Mooij2014,Stegle2010}, infer interactions in terms of information flow in a probabilistic framework and cannot detect bidirectional causality.
A recent suite of methods based on dynamical systems theory proposes to infer interactions, both unidirectional and bidirectional, in systems that are nonlinear, noisy, and potentially high-dimensional \cite{Sugihara2012, Ye2015, Clark2015}.
The basic idea is that if $X$ drives $Y$, information about $X$ is embedded in the time series of $Y$.
Examining the relationships between delay-embeddings of the time series of $X$ and $Y$ can reveal whether $X$ drives $Y$, $Y$ drives $X$, both, or neither.
These approaches, which we refer to collectively as convergent cross-mapping (CCM), have been offered as general tools to analyze causation in nonlinear dynamical systems \cite{Sugihara2012, Ye2015, Clark2015}.

\begin{figure}
\begin{center}
\includegraphics[width=6in]{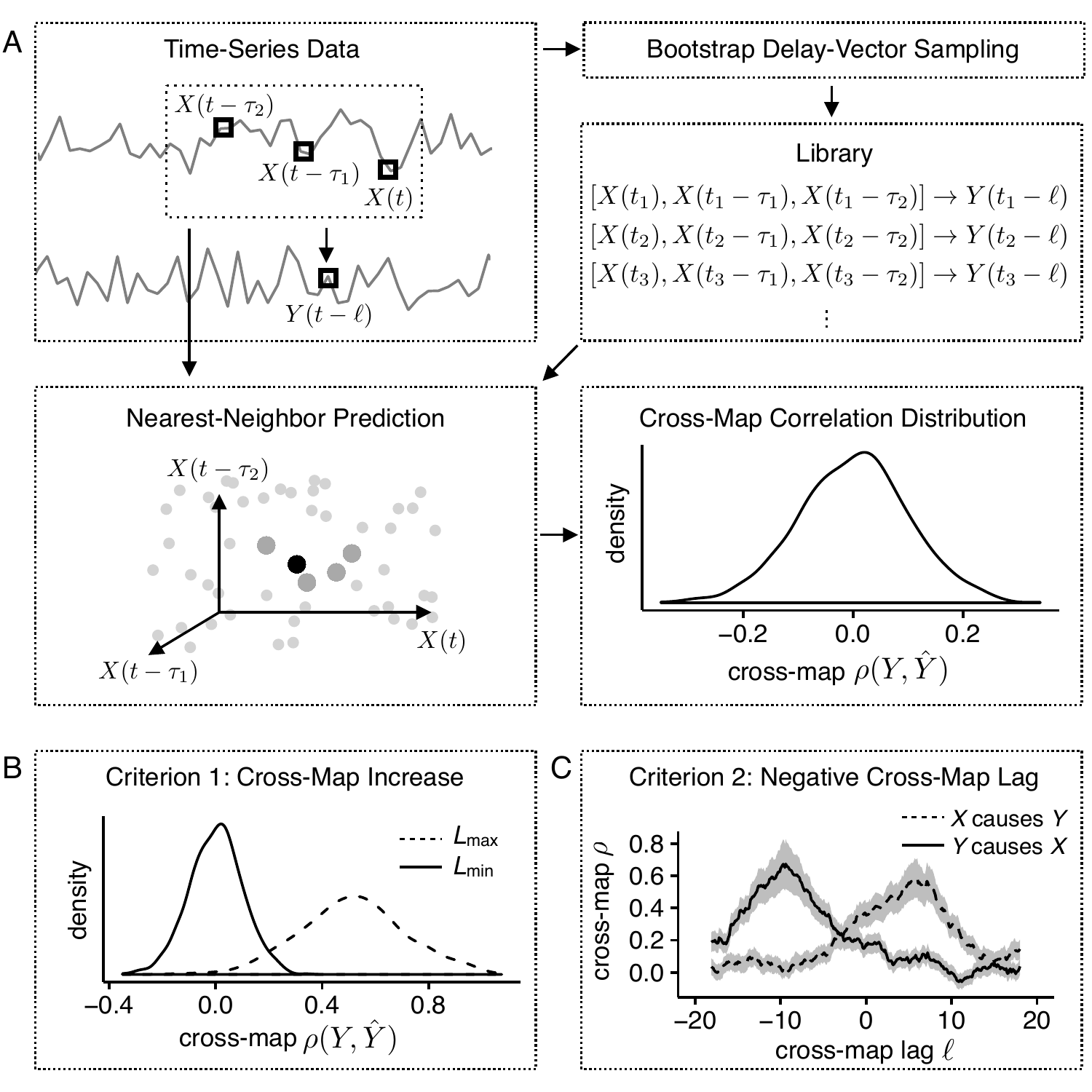}
\end{center}
\caption{\textbf{Summary of criteria for detecting causality}. (A) Schematic of cross-map algorithm for testing $Y \rightarrow X$. Delay vectors in $X$, mapped to values in $Y$ with lag $\ell$, are bootstrap-sampled to construct a prediction library. For each delay vector in $X$, reconstructed values $\hat{Y}$ are calculated from a distance-weighted sum of $Y$ values from nearest neighbors in the library. Many sampled libraries yield a distribution of cross-map correlations between actual $Y$ and reconstructed $\hat{Y}$. (B) Criterion 1 (cross-map increase). Bootstrap distributions of cross-map correlation are calculated at minimum and maximum library sizes with $\ell = 0$; causality is inferred if the correlation at $L_{\max}$ is significantly greater than the correlation at $L_{\min}$. (C) Criterion 2 (negative cross-map lag). Cross-map correlations are calculated across different values of $\ell$. Causality is inferred if the highest cross-map correlation for negative $\ell$ is positive and significantly greater than the highest value for nonnegative $\ell$. \label{fig:conceptual}}
\end{figure}

The mathematical foundations of CCM, and therefore its assumptions, lie in deterministic nonlinear systems theory.
After sufficient time, the states of a deterministic dynamical system reach an attractor, which may be a point equilibrium, a limit cycle, or a higher-dimensional chaotic attractor.
By Takens' theorem, a one-dimensional time series $X(t)$ from the system can be mapped perfectly to the attractor in the full state space in the system by constructing a delay embedding, in which states of the full system are mapped to delay vectors, $\bx(t) = \{X(t), X(t - \tau_1), X(t - \tau_2), \ldots, X(t - \tau_{E-1} \}$, for \emph{delays} $\tau_i$ and an \emph{embedding dimension} $E$, which must be at least as large as the dimensionality of the attractor~\cite{Takens1981}.
This mapping provides the basis for causal inference under CCM: if $Y$ drives (causes) $X$, then a newly observed $\bx(t)$ can perfectly reconstruct the corresponding $\hY(t)$ from past observations of the mapping $\bx(t) \rightarrow Y(t)$ (Fig.~\ref{fig:conceptual}A).
As the number of observed delay vectors $\bx(t)$ increases, the reconstruction converges to small error, as observed points on the reconstructed attractor become close together~\cite{Sugihara2012}.

With finite, noisy real data, the reconstruction is necessarily imperfect, and two operational criteria have been used to detect causality.
The first criterion (Fig.~\ref{fig:conceptual}B) is based simply on this improvement in reconstruction quality with the number of observations.
This approach is known to produce false positives in the case of strongly driven variables, where the system becomes synchronized to the driver \cite{Sugihara2012, Kocarev1996}.
This failure is logically consistent with the theory: the theory implies that, with perfect data, causal drivers will produce good reconstructions, but not that non-causal drivers will not produce good reconstructions.
The second criterion (Fig.~\ref{fig:conceptual}C) tries to correct this problem by additionally considering the directionality of information flow in time~\cite{Ye2015}.
If one variable drives another, the best predictions of current states of the driven variable should come from past, not current or future, states of the driver.

Many ecological systems undergo synchronized diurnal or annual fluctuations and thus raise doubts about the first criterion.
Transient dynamics, demographic and environmental noise, and observation error---all ubiquitous in nature---raise general concerns, since they violate the theory's assumption that variables are perfectly observed in a deterministic system.
Variations of CCM have nonetheless been applied to such systems to test hypotheses about who interacts with whom \cite{Ye2015, Sugihara2012, Tajima2015, Tsonis2015, Clark2015}.

We investigated whether the frequently periodic, noisy, and transient dynamics of ecological systems are a current obstacle to causal inference based on state-space reconstruction.
These factors have been addressed to varying degrees in different contexts \cite{Sugihara2012, Ye2015, Clark2015} but not systematically.
Specifically, we examined whether the two criteria for causal inference are robust to inevitable uncertainties about the dynamics underlying the data.
With little prior knowledge of a system's complexity, including the influences of transient dynamics and noise, can we reach statistically rigorous conclusions about who interacts with whom?
Infectious diseases provide a useful test case because their dynamics have been extensively studied, long time series are available, and pathogens display diverse immune-mediated interactions \cite{Cobey2014}.
Their dynamics are also influenced by seasonal variation in transmission rates, host population structure, and pathogen evolution.
The ability to test directly for the presence of interactions would save considerable effort over fitting semi-mechanistic models that incorporate these complexities.
We find that although CCM appears to work beautifully in some instances, it does not in others.
Noise and transient dynamics contribute to poor outcomes, as do statistical ambiguities in the methodology itself.
We propose that except in extreme circumstances, the current method cannot reliably reveal causality in natural systems.

\section*{Results}

To assess the reliability of CCM, we began by simulating the dynamics of two strains with stochastic, seasonally varying transmission rates (Methods).
In large systems, many factors might influence these rates.
In low-dimensional models, these factors are typically represented as process noise.
We consequently varied the level of process noise in our simulations by changing its standard deviation, $\eta$.
We also varied the strength of competition from strain 2 on strain 1 ($\sigma_{12}$); strain 1, in contrast, never affected strain 2 ($\sigma_{21}=0$).
For each level of competition and process noise, we simulated 100 replicates from random initial conditions to stochastic fluctuations around a deterministic attractor.
One thousand years of error-free monthly incidence were output to give CCM the best chance to work.
For each combination of parameters (competition strength $\sigma_{12}$ and process noise $\eta$), we examined whether strain interactions were correctly inferred.
When $\sigma_{12} >0$, strain 2 should be inferred to ``drive'' (influence) strain 1.
Because $\sigma_{21}=0$, strain 1 should never be inferred to drive strain 2.

To detect interactions, for each individual time series, we identified the delay-embeddings (Fig. \ref{fig:conceptual}A) and applied one of two causality criteria using the reconstructed attractors (Fig. \ref{fig:conceptual}B,C and Methods).
Both criteria are based on the cross-map correlation $\rho$, which is the correlation between reconstructed values of $\hat{Y}$ and actual values of $Y$, given the reconstructed attractor of $X$.
We use $p < 0.05$ to identify significant differences in these correlations because we are interested in situations in which the null hypothesis of no change in correlation, and thus no interaction, is rejected.
Criterion 1 \cite{Sugihara2012, Clark2015} measures whether the cross-map correlation increases as the number of observations of the putatively driven variable grows (Fig. \ref{fig:conceptual}B).
We refer to this as the cross-map increase criterion.
Criterion 2 \cite{Ye2015} infers a causal interaction if the maximum cross-correlation of the putative driver is positive and occurs in the past (i.e., at a negative temporal lag; Fig. \ref{fig:conceptual}C).
We refer to this as the negative cross-map lag criterion.
For simplicity, we start with Criterion 1.

\subsection*{Sensitivity to periodicity}

\begin{figure}
\begin{center}
\includegraphics[width=6in]{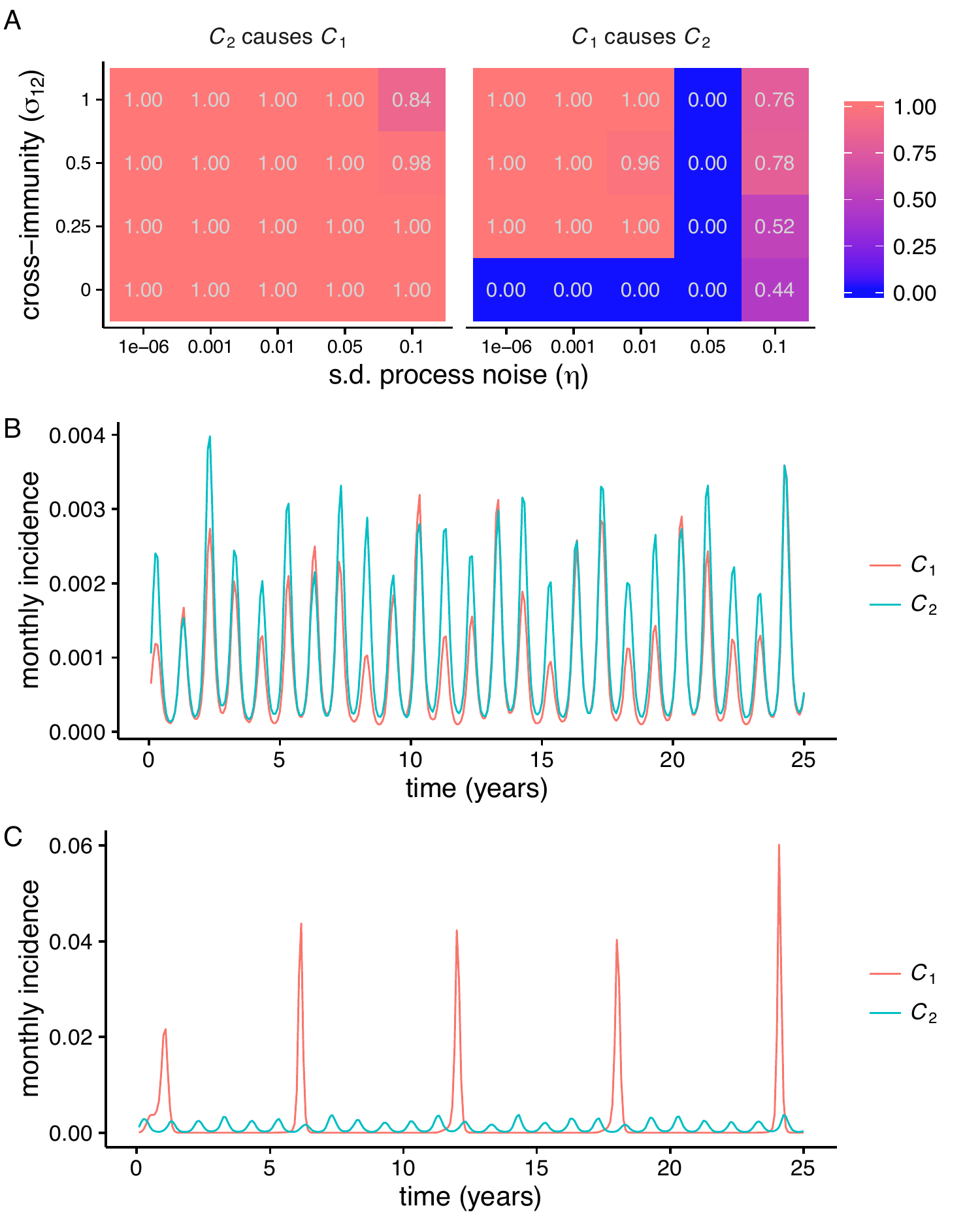}
\end{center}
\caption{\textbf{Interactions detected as a function of process noise and the strength of interaction ($C_2 \rightarrow C_1$) and representative time series}. (A) Heat maps show the fraction of 100 replicates significant for each inferred interaction for different parameter combinations. A significant increase in cross-map correlation $\rho$ with library length $L$ indicated a causal interaction. The time series consisted of 1000 years of monthly data. (B) Representative 25-year sample of the time series for which mutual interactions were inferred ($\sigma_{12}=0.25, \eta=0.01$). (C) Representative sample of the time series for which $C_2$ is inferred to drive $C_1$ but not vice-versa ($\sigma_{12}=0.25, \eta=0.05$).  \label{fig:univar_monthly_hm_tmp}}
\end{figure}

Criterion 1, which requires a significant increase in cross-map correlation $\rho$ with observation library size $L$, frequently detected interactions that did not exist.
In all cases where strain 2 had no effect on strain 1, CCM always incorrectly inferred an influence (Fig.~\ref{fig:univar_monthly_hm_tmp}A).
Although strain 1 never influenced strain 2, it was often predicted to (Fig.~\ref{fig:univar_monthly_hm_tmp}A).
Sample time series suggested a strong correlation between synchronous oscillations and the appearance of bidirectional interactions (Fig.~\ref{fig:univar_monthly_hm_tmp}B).
In contrast, when strain 2 appeared to drive strain 1 but not vice-versa ($\sigma_{12}=0$ and $\eta=0.05$), strain 1 often oscillated with a period that was an integer multiple of the other strain's (Fig.~\ref{fig:univar_monthly_hm_tmp}C).
Thus, as expected, strongly synchronized dynamics prevented separation of the variables.
Additionally, the resemblance of strain 2 to the seasonal driver led to false positives even when the strains were independent and strain 1 oscillated at a different frequency.

\begin{figure}
\begin{center}
\includegraphics[width=6in]{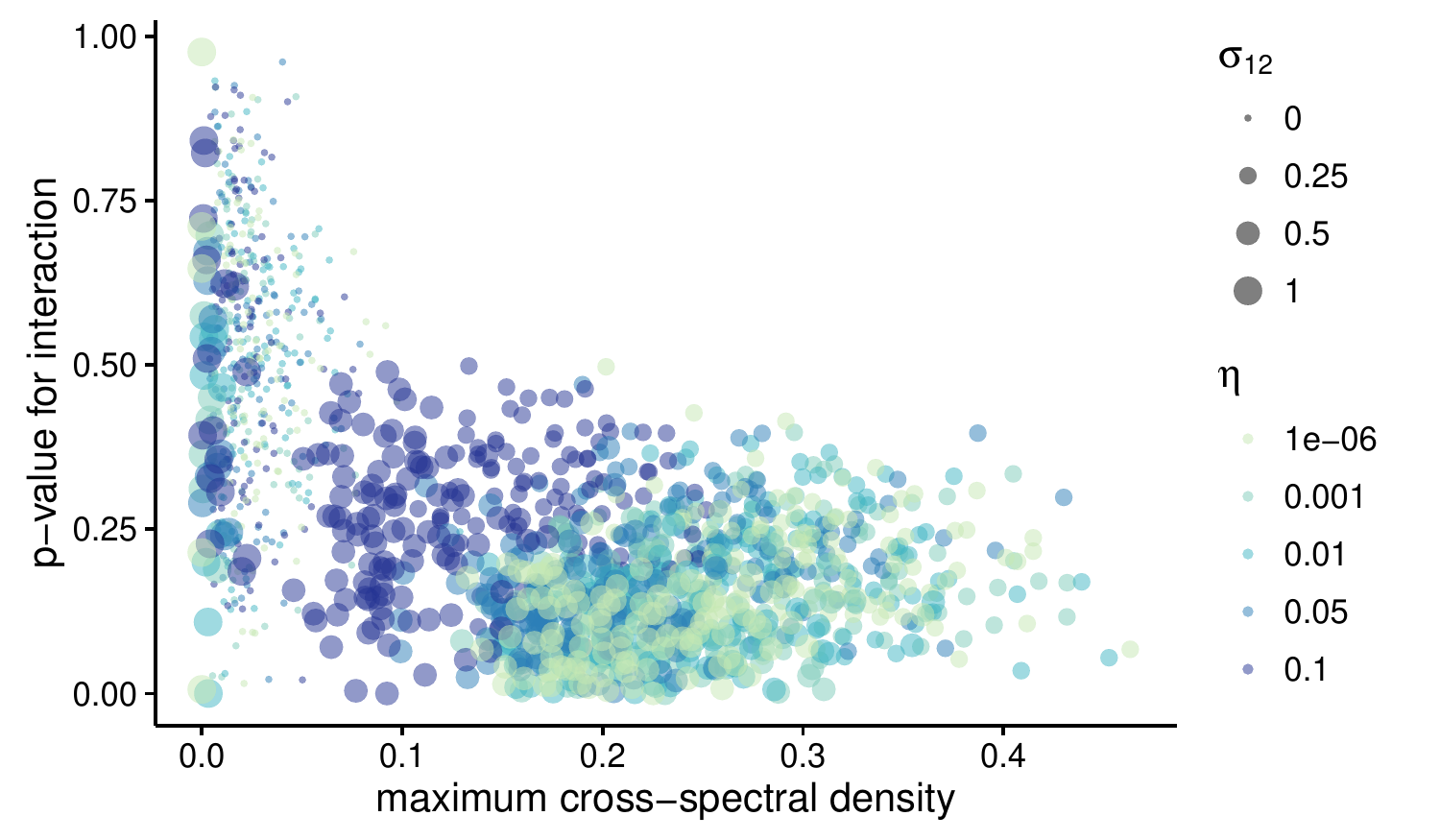}
\end{center}
\caption{\textbf{Shared frequency spectra predict probability of inferred interaction}. Points show the maximum cross-spectral densities of strains 1 and 2 plotted against the p-values for $C_1 \rightarrow C_2$ for 1000 years of annual data. In all replicates, $C_1$ never actually drives $C_2$. Point color indicates the strength of $C_2 \rightarrow C_1$ ($\sigma_{12}$), and point size indicates the standard deviation of the process noise ($\eta$) on transmission rates. \label{fig:max_crossspec_tmp}}
\end{figure}

\begin{figure}
\begin{center}
\includegraphics[width=6in]{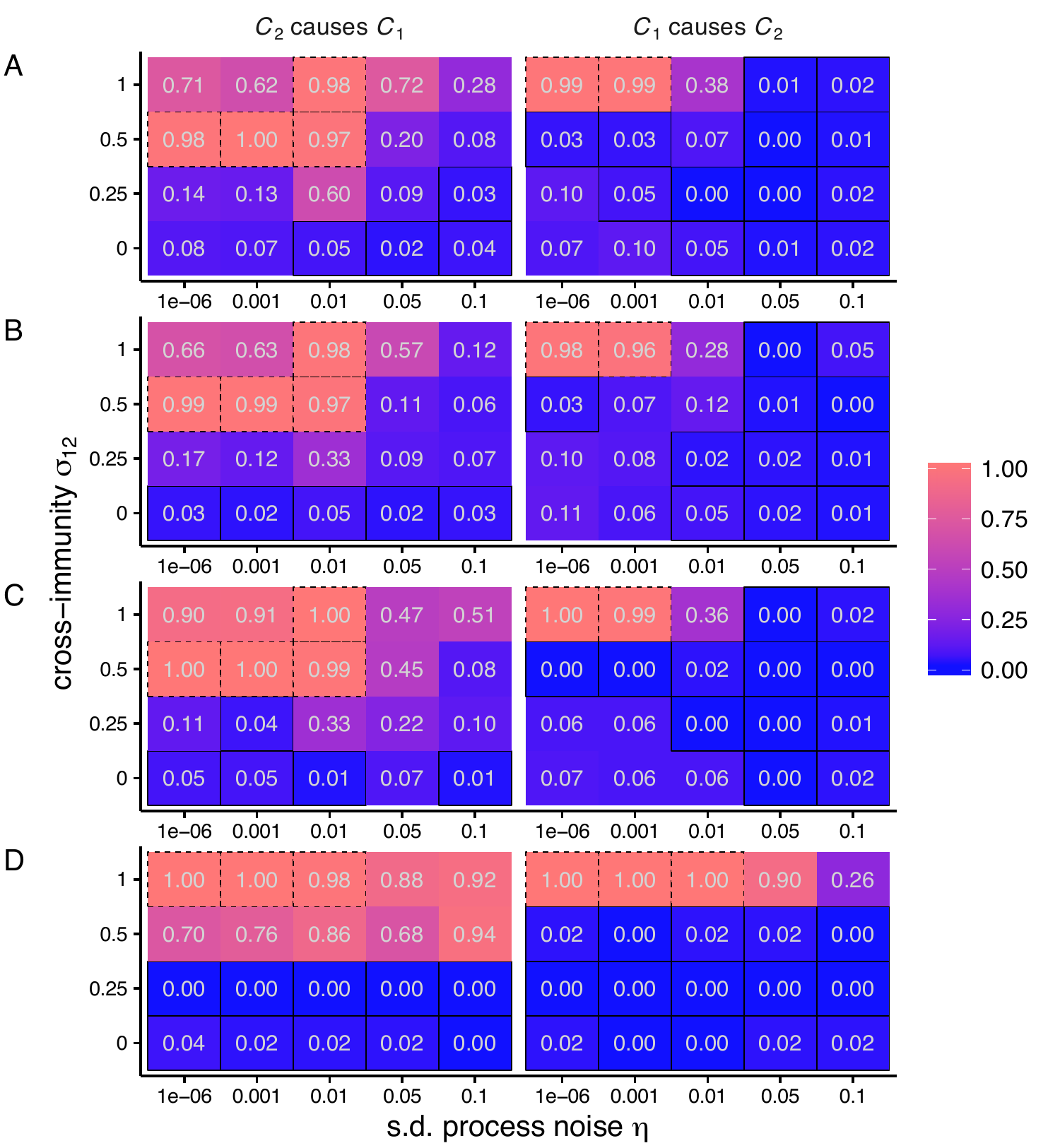}
\end{center}
\caption{\textbf{Interactions detected as a function of process noise and the strength of interaction ($C_2 \rightarrow C_1$) for different types of data}. Heat maps show the fraction of 100 replicates significant for each inferred interaction for different parameter combinations. A significant increase in cross-map correlation $\rho$ with library length $L$ indicated a causal interaction. Each analysis is based on 1000 years of data. (A) Annual incidence, (B) prevalence strobed annually, (C) first-differenced annual incidence, and (D) monthly incidence without seasonal forcing.  \label{fig:detect_diffdata}}
\end{figure}

The sensitivity of the method to periodicity persisted despite transformations of the data and changes to the  driver.
One possible solution to reducing seasonal effects, sampling annual rather than monthly incidence, reduced the overall rate of false positives but also failed to detect some interactions (Fig.~\ref{fig:detect_diffdata}A).
Furthermore, when the effects of strain 2 on 1 were strongest, the reverse interaction was more often inferred.
Sampling the prevalence at annual intervals gave similar results (Fig.~\ref{fig:detect_diffdata}B), and first-differencing the data did not qualitatively change outcomes (Fig.~\ref{fig:detect_diffdata}C).
The method yielded incorrect results even without seasonal forcing ($\epsilon=0$) because of noise-induced oscillations (Fig.~\ref{fig:detect_diffdata}D).
In all of these cases, the presence of shared periods between the strains correlated strongly and significantly with the rate of detecting a false interaction (Fig.~\ref{fig:max_crossspec_tmp}).

Because cross-map skill should depend on the quality of the reconstructed attractor, we investigated performance under other methods of constructing the attractors of the two strains (Methods).
Nonuniform embedding methods allow the time delays to occur at irregular intervals, $\tau_1, \tau_2, ... \tau_{E-1}$, which may provide a more accurate reconstruction.
Alternative reconstruction methods, including nonuniform embedding \cite{Nichkawde2013, Uzal2011}, random projection \cite{Tajima2015}, and maximizing the cross-map (rather than univariate) correlation failed to fix the problem (Fig.~\ref{fig:detect_diffembed}).

\begin{figure}
\begin{center}
\includegraphics[width=6in]{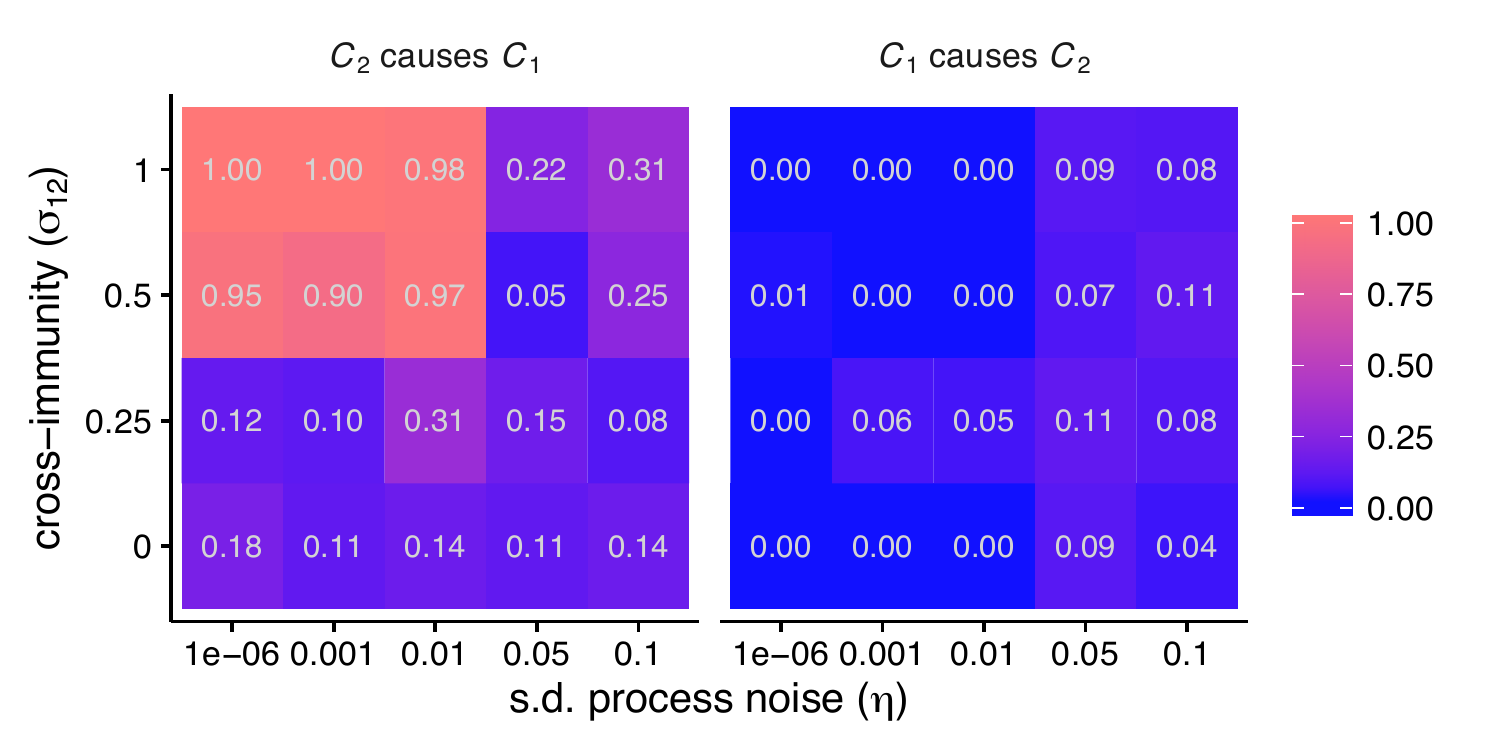}
\end{center}
\caption{\textbf{Interactions detected as a function of process noise and the strength of interaction ($C_2 \rightarrow C_1$) and representative time series}. Heat maps show the fraction of 100 replicates significant for each inferred interaction for different parameter combinations. A maximum, positive cross-map correlation $\rho$ at a negative lag indicated a causal interaction. Each replicate used 100 years of monthly incidence. \label{fig:univar_monthly_lag_seas_diff_tmp}}
\end{figure}

\begin{figure}
\begin{center}
\includegraphics[width=6in]{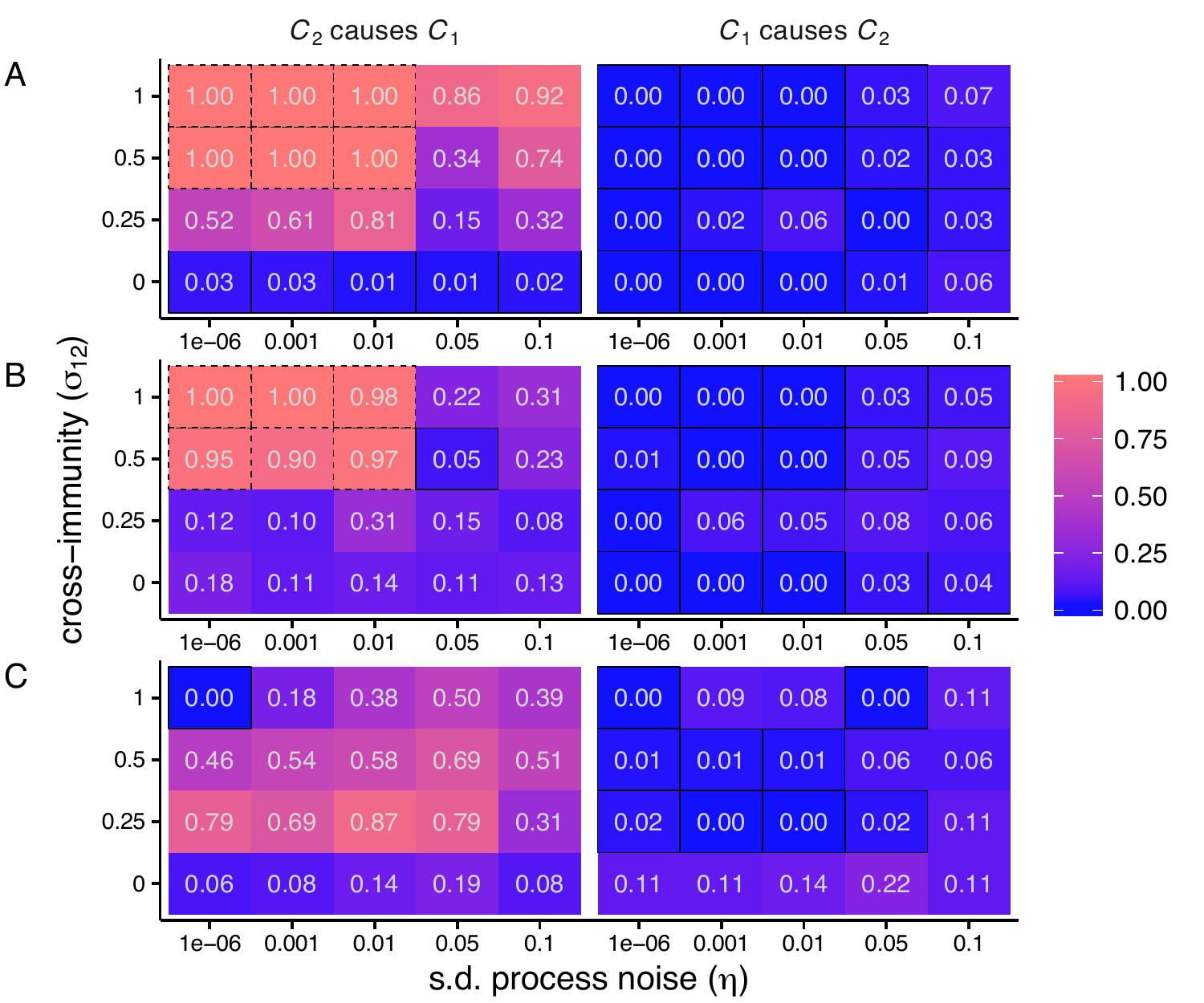}
\end{center}
\caption{\textbf{Interactions detected for different types of data}. Heat maps show the fraction of 100 replicates significant for each inferred interaction for different parameter combinations. A maximum cross-map correlation $\rho$ at a negative lag was required for inferring causal interaction. (A) 1000 years of annual incidence, requiring that the maximum $\rho$ be positive. (B) 100 years of monthly incidence, requiring that the maximum $\rho$ be increasing. (C) 100 years of monthly incidence with identical strains ($\beta_1=\beta_2=0.3$), requiring that maximum $\rho$ be positive.  \label{fig:detect_diffdata_lag}}
\end{figure}

Criterion 2, which infers that $Y$ drives $X$ if there is a positive cross-map correlation that is maximized at a negative cross-map lag, performed relatively well (Fig.~\ref{fig:univar_monthly_lag_seas_diff_tmp}).
Fewer false positives were detected, although the method missed some weak extant interactions ($\sigma_{12}=0.25$) and interactions in noisy systems ($\eta=0.05, 0.1$).
Results for annual data were similar (Fig.~\ref{fig:detect_diffdata_lag}A).
Requiring that $\rho$ be not only positive but also increasing barely affected performance (Fig.~\ref{fig:detect_diffdata_lag}B).

\subsection*{Limits to identifiability}

If two variables $X$ and $Y$ share the same driver but do not interact, if the driving is strong enough, $X$ may resemble the driver so closely that $X$ appears to drive $Y$.
In a similar vein, when the two strains in our system have identical transmission rates ($\beta_1= \beta_2$) and one strongly drives the other ($\sigma_{12}=1$), the direction of the interaction cannot be detected when the dynamics are nearly deterministic ($\eta=10^{-6}$) (Fig.~\ref{fig:detect_diffdata_lag}C).
Causal inference in such cases becomes difficult.

\begin{figure}
\begin{center}
\includegraphics[width=4in]{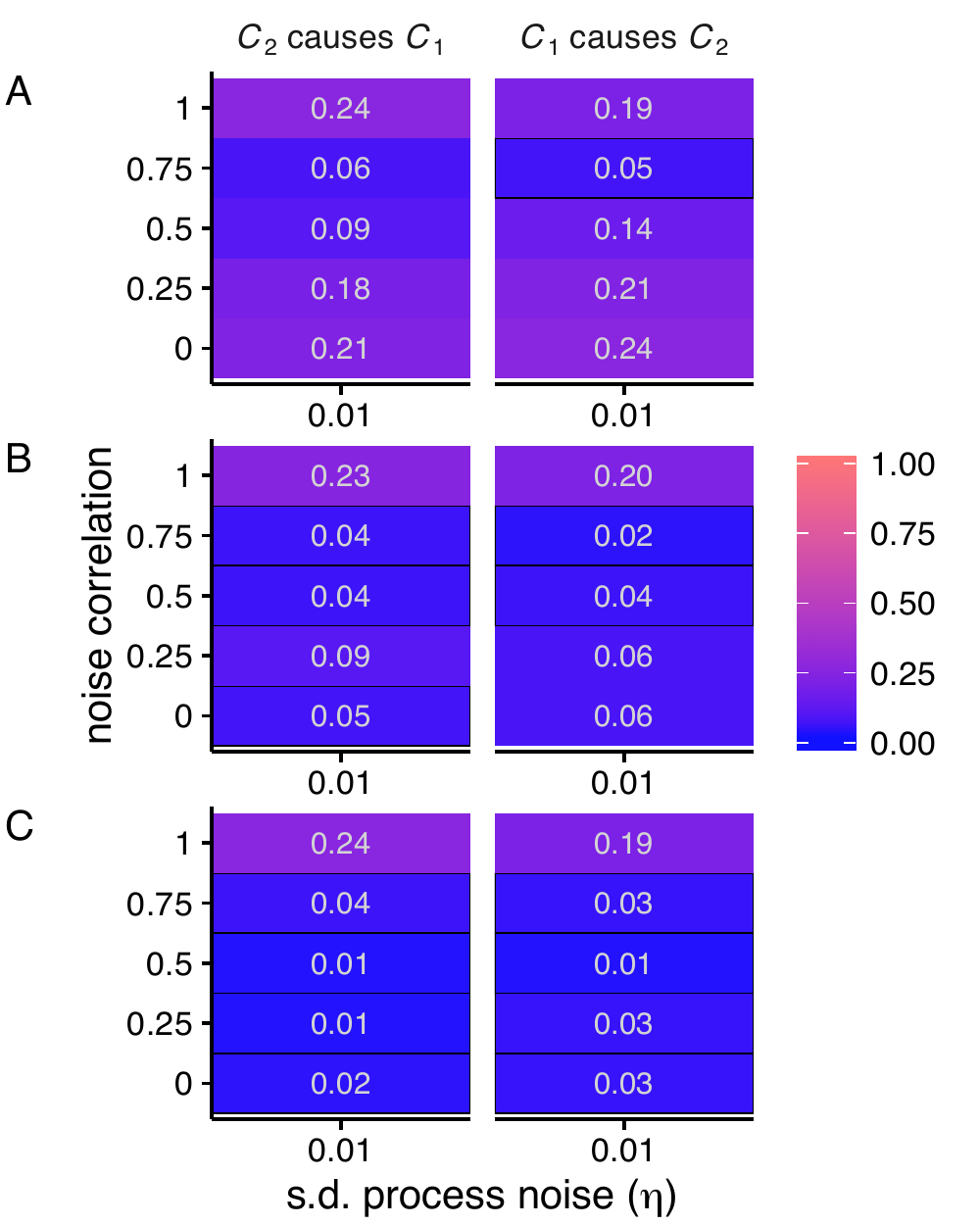}
\end{center}
\caption{\textbf{Interactions detected between identical strains with correlated process noise}. Heat maps show the fraction of 100 replicates significant for each inferred interaction. A maximum cross-map correlation $\rho$ at a negative lag was required for inferring causal interaction. 100 years of monthly (A) and 1000 years of annual (B) incidence, requiring that the maximum $\rho$ be positive. (C) 100 years of monthly incidence, requiring that maximum $\rho$ be increasing.  \label{fig:detect_corrproc_identical}}
\end{figure}

To investigate the limits to distinguishing strains that are ecologically similar and do not interact, we varied the correlation of the strain-specific process noise while applying the more conservative of the two criteria for inferring causality (Criterion 2), that the cross-map correlation $\rho$ be positive and peak at a negative lag \cite{Ye2015}.
Process noise can be thought of as a hidden environmental driver that affects both strains simultaneously, and thus the strength of correlation indicates the relative contribution of shared versus strain-specific noise.
With two identical, independent strains, no seasonal forcing, and low process noise ($\eta=0.01$), the false positive rate depended on correlation strength and the quantity of data.
When using 100 years of monthly incidence, the false positive rate varied non-monotonically with correlation strength, with a minimum (5\%-6\%) at a correlation of 0.75 and its highest values, near 24\%, at correlations of 0 and 1 (Fig.~\ref{fig:detect_corrproc_identical}A).
Using 1000 years of annual incidence reduced false positive rates to 5\%-9\% for imperfectly correlated noise (Fig.~\ref{fig:detect_corrproc_identical}B).
The best performance occurred with 100-year monthly data when cross-map correlation was required to increase with library length (Fig.~\ref{fig:detect_corrproc_identical}C).
Thus, the independence of two strains will generally be detected as long as they experience imperfectly correlated noise.

\begin{figure}
\begin{center}
\includegraphics[width=6in]{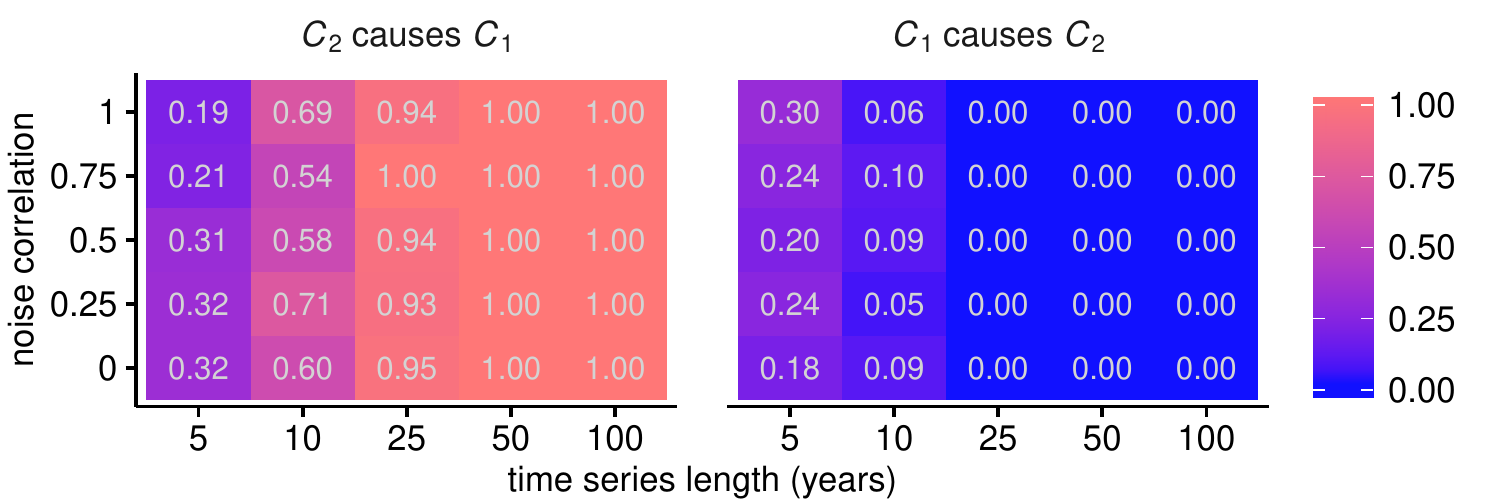}
\end{center}
\caption{\textbf{Interactions detected between distinct strains with correlated process noise}. Heat maps show the fraction of 100 replicates significant for each inferred interaction. A maximum cross-map correlation $\rho$ at a negative lag and $\rho>0$ were required for inferring causal interaction. Results are shown for 5, 10, 25, 50, and 100 years of monthly incidence.  \label{fig:detect_corrproc_distinct}}
\end{figure}

We next considered the problem of identifying two ecologically distinct strains ($\beta_1 \neq \beta_2$) when one strain strongly drives the other ($\sigma_{12}=1$) and its dynamics resemble the seasonal driver.
In this case, even with perfectly correlated process noise, correct interactions are consistently inferred (Fig.~\ref{fig:detect_corrproc_distinct}).
Thus, we conclude that the presence of noise, even highly correlated noise, can help distinguish causality between coupled, synchronized variables \cite{Schumacher2015}.
It is more difficult to distinguish non-interacting, dynamically equivalent variables.
In the latter case, noise has inconsistent effects on causal inference, although Criterion 2 may perform much better than Criterion 1.
These results at least hold for ``modest'' noise ($\eta=0.01$): as shown earlier, higher levels hurt performance (Fig.~\ref{fig:univar_monthly_lag_seas_diff_tmp}).

\subsection*{Transient dynamics}
CCM is optimized for dynamics that have converged to a deterministic attractor.
Directional parameter changes in time and large perturbations can prevent effective cross-mapping because the method requires a consistent mapping between system states as well as sufficient coverage of state space by the data.
We evaluated the impact of both of these types of transient dynamics on causal inference, using a simple example of each as proof of principle.

\begin{figure}
\begin{center}
\includegraphics[width=6in]{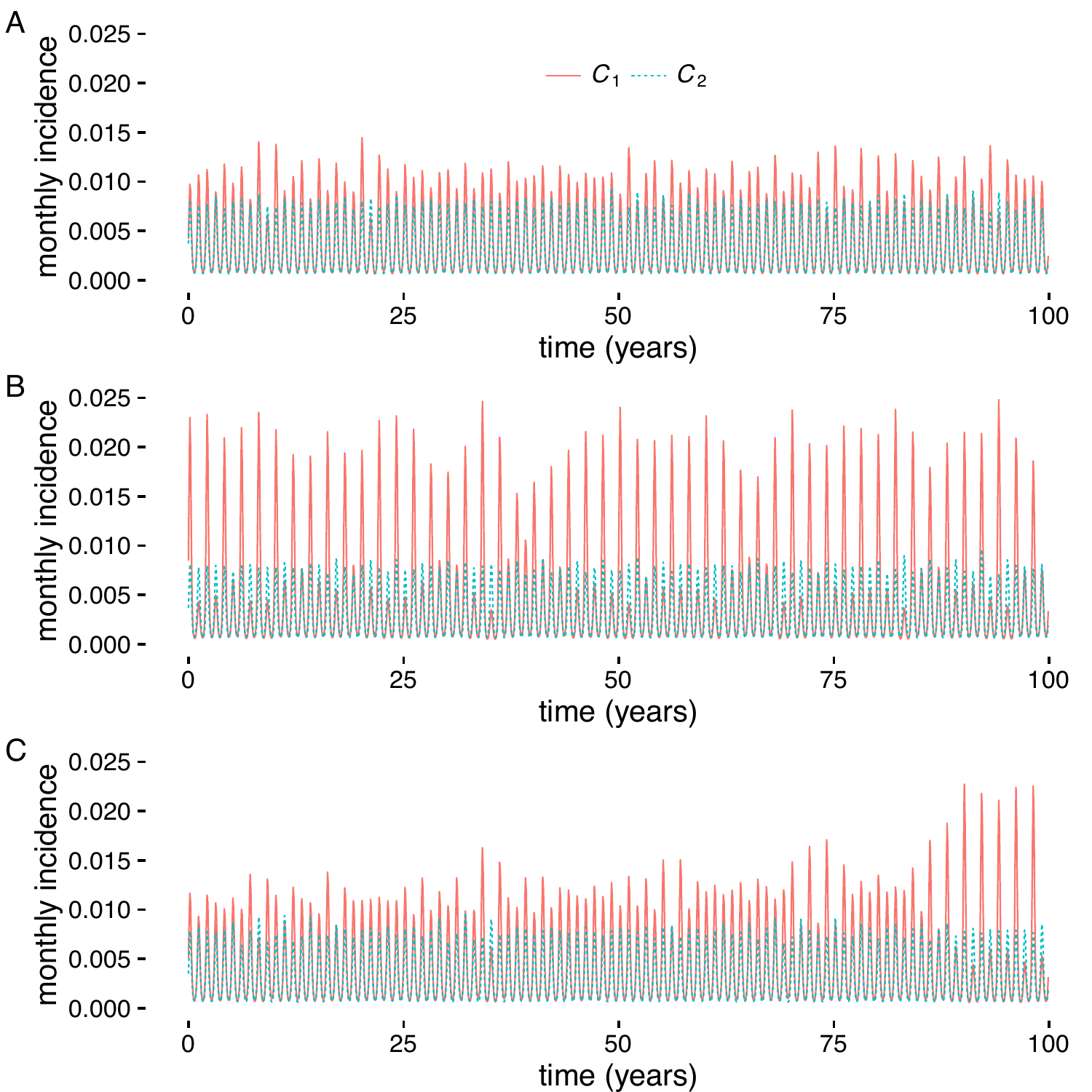}
\end{center}
\caption{\textbf{Incorrect inference with changing transmission rate.} Example time series for testing transient dynamics. Each time series contained 100 years of monthly incidence data. The transmission rate $\beta_1$ for the driven strain $C_1$ was fixed at $\beta_1 = 0.30$ (A) and $\beta_1 = 0.32$ (B), and varied linearly over time between the two values (C). The transient time series yields high false positive and false negative rates under CCM. Interaction strength was $\sigma_{21} = 0.5$, process noise was $\eta = 0.01$, and seasonal forcing was $\epsilon = 0$. \label{fig:betachange}}
\end{figure}

In the first test, we identified two sets of parameter values where CCM was successful under Criterion 2 (intermediate interaction strength, $\sigma_{21} = 0.5$; seasonal forcing, $\epsilon = 0.1$; process noise, $\eta = 0.01$; and transmission rates $\beta_1$ of $0.30$ (Fig.~\ref{fig:betachange}A) and $0.32$ (Fig.~\ref{fig:betachange}B)).
We tested CCM on simulations with the parameter values fixed and then with the transmission rate $\beta_1$ varying linearly over time betwen the two values.
All three tests used 100 years of monthly incidence.
Of 100 replicates, with $\beta_1$ fixed at $0.30$, CCM failed to detect an interaction 5 times, and never falsely detected an absent interaction.
With $\beta1$ fixed at $0.32$, there were 12 false negatives and 1 false positive.
When $\beta_1$ varied from $0.30$ to $0.32$, error rates increased: there were 29 false negatives and 44 false positives.
Transient dynamics due to a linear change in a system parameter can thus lead to incorrect causal inference even when causal inference is successful before and after the change.

\begin{figure}
\begin{center}
\includegraphics[width=6in]{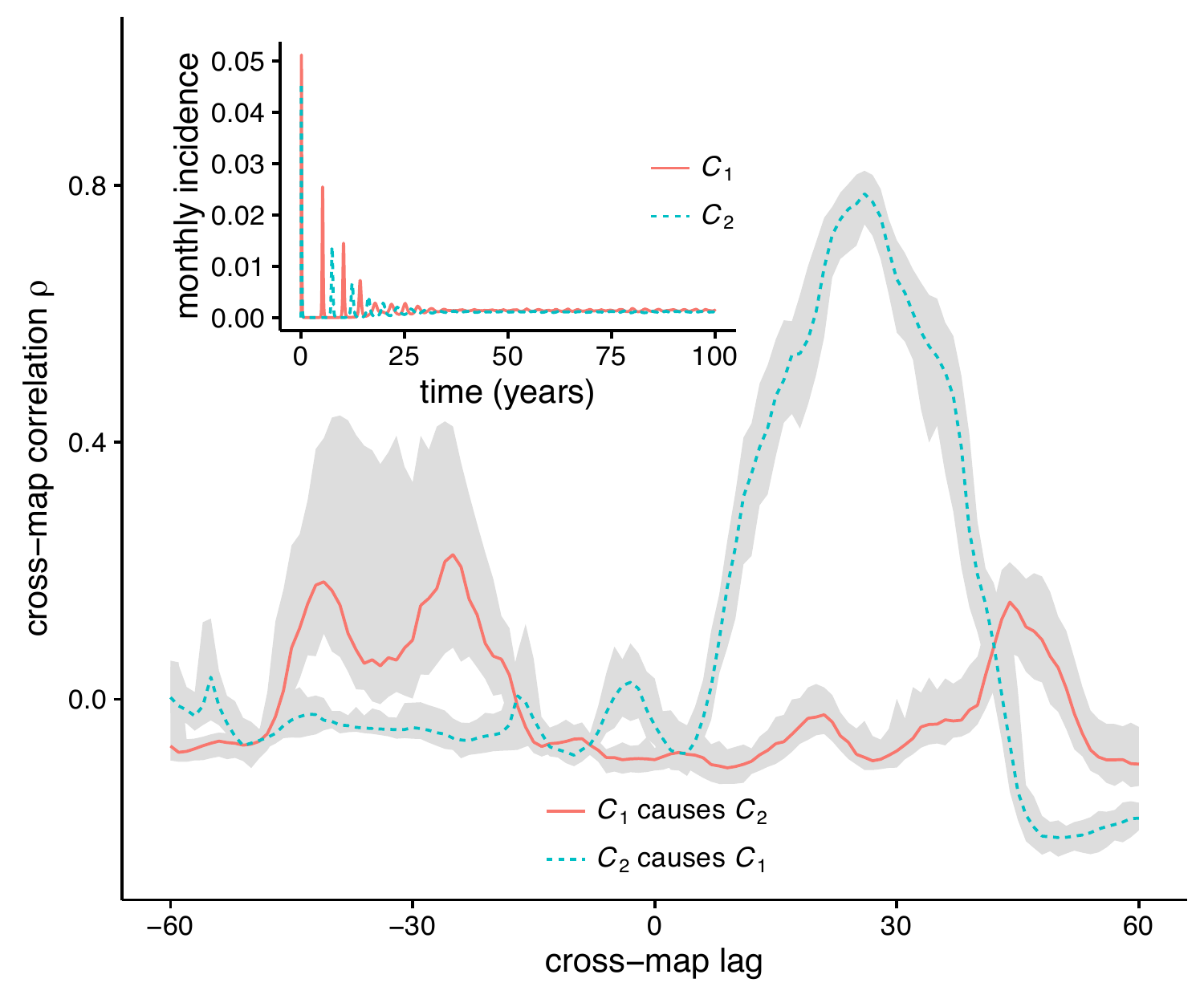}
\end{center}
\caption{\textbf{Incorrect inference with far-from-attractor dynamics.} Cross-map correlations at different lags for a sample 100-year time series with monthly sampling (inset). Lines represent bootstrap medians; gray ribbons represent the middle 95\% of the bootstrap distribution. Although $C_2$ drives $C_1$ ($\sigma_{12}= 0.5, \sigma_{21}=0$), the maximum cross-correlation $\rho$ for $C_1$ cross-mapped to $C_2$ occurs at a positive lag, and the reverse at a negative lag, leading to the conclusion that $C_1$ drives $C_2$, and $C_2$ does not drive $C_1$. Sample dynamics include process noise ($\eta=0.01$) but no seasonal forcing ($\epsilon=0$). \label{fig:transient}}
\end{figure}

In the second test, we began simulations at random initial conditions far from equilibrium and applied CCM to the first 100 years of monthly incidence.
When strain 2 weakly drives strain 1 ($\sigma_{12}=0.5$), causal inference is compromised, even when process noise is low ($\eta=0.01$; Fig.~\ref{fig:transient}).
In 100 simulations of this scenario, the correct interaction (strain 2 driving strain 1) was always detected after transients had passed, but it was detected in only 19 of 100 simulations that included transients.
Furthermore, a reverse interaction (strain 1 driving 2) was incorrectly detected in 21 of 100 simulations.
The method thus performed worse than chance in identifying interactions that were present, and it also regularly predicted nonexistent interactions.

\subsection*{Application to childhood infections}

\begin{figure}
\begin{center}
\includegraphics[width=6in]{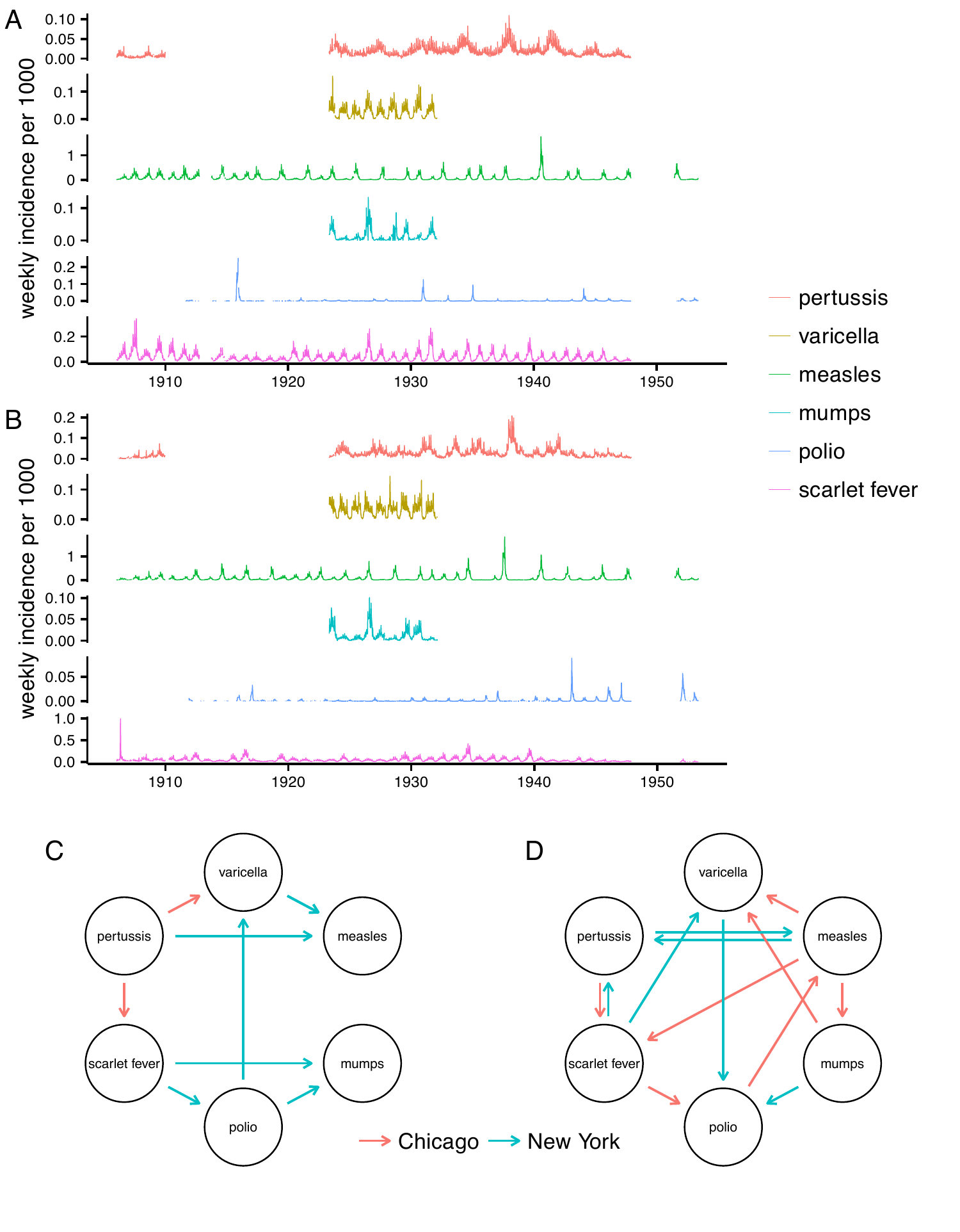}
\end{center}
\caption{\textbf{Historical childhood infections in New York City and Chicago and inferred interactions from two reconstruction methods.} Time series show weekly incidence of infections per 1000 inhabitants of New York City (A) and Chicago (B). Delay-embeddings were constructed by maximizing the univariate correlation (C) or through a random projection method (D) Arrows indicate the inferred interactions from the New York (blue) and Chicago (red) time series under Criterion 2 (negative cross-map lag).\label{fig:historical_data_tmp}}
\end{figure}

\begin{figure}
\begin{center}
\includegraphics[width=6in]{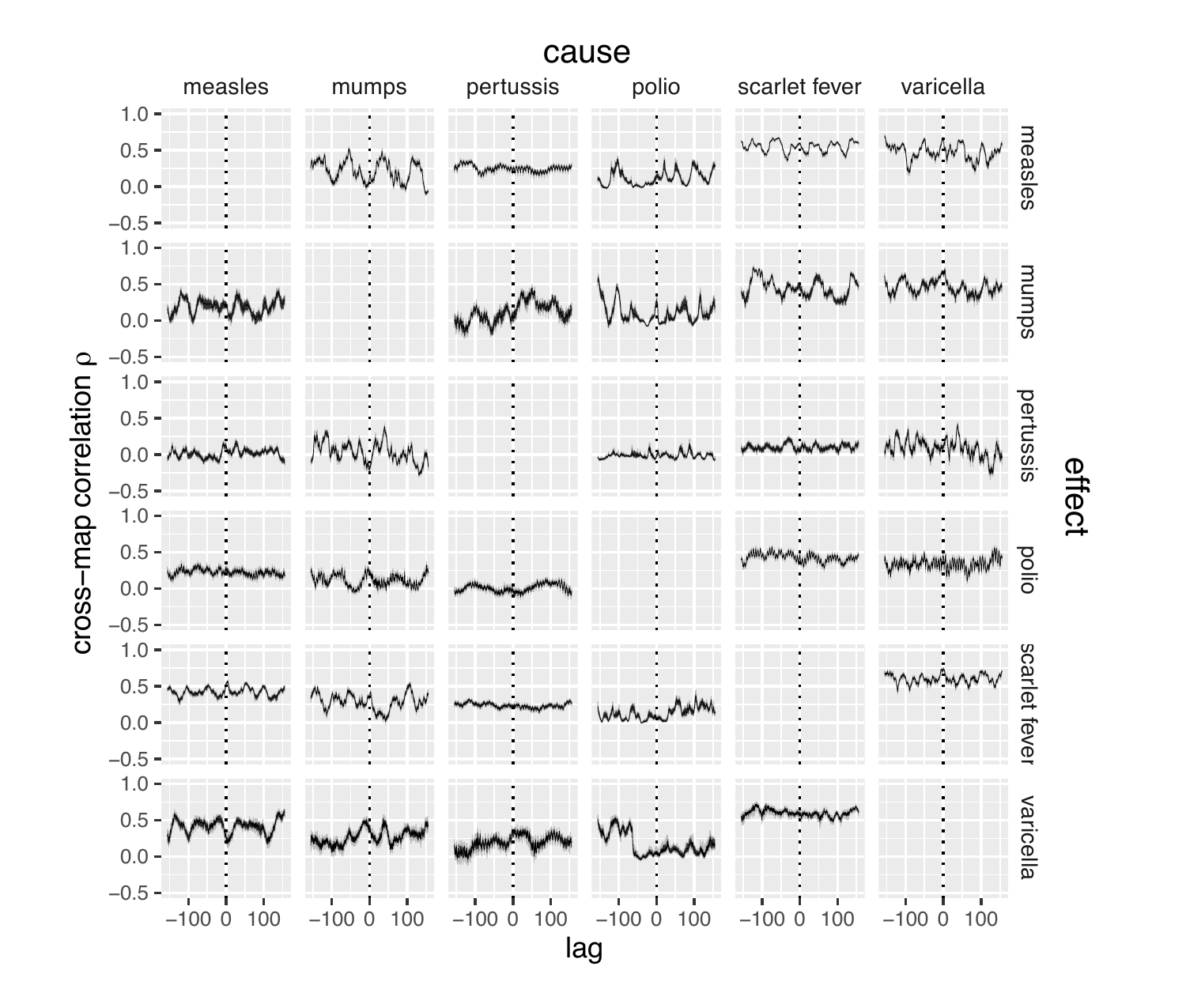}
\end{center}
\caption{\textbf{Cross-map lags for New York with default (univariate) embedding}.  \label{fig:cities_corrbylag_nyc_self_uniform}}
\end{figure}

Given the apparent success of CCM under Criterion 2 (negative cross-map lag) with two strains and little noise near the attractor, we investigated whether the method might shed light on the historic dynamics of childhood infections in the pre-vaccine era.
Time series analyses have suggested that historically common childhood pathogens may have competed with or facilitated one another \cite{Mina2015, Rohani2003}.
We obtained the weekly incidence of six reportable infections in New York City from intermittent periods spanning 1906 to 1953 \cite{vanPanhuis2013} (Fig.~\ref{fig:historical_data_tmp}A).
Six of 30 pairwise interactions were significant at the $p<0.05$ level, not correcting for multiple tests (Fig.~\ref{fig:historical_data_tmp}C).
Polio drove mumps and varicella, scarlet fever drove mumps and polio, and varicella and pertussis drove measles.
Typical cross-map lags occurred at one to three years (Fig.~\ref{fig:cities_corrbylag_nyc_self_uniform}).
The inferred interactions were identical if we required that the cross-map correlation $\rho$ be increasing and not merely positive.

\begin{figure}
\begin{center}
\includegraphics[width=6in]{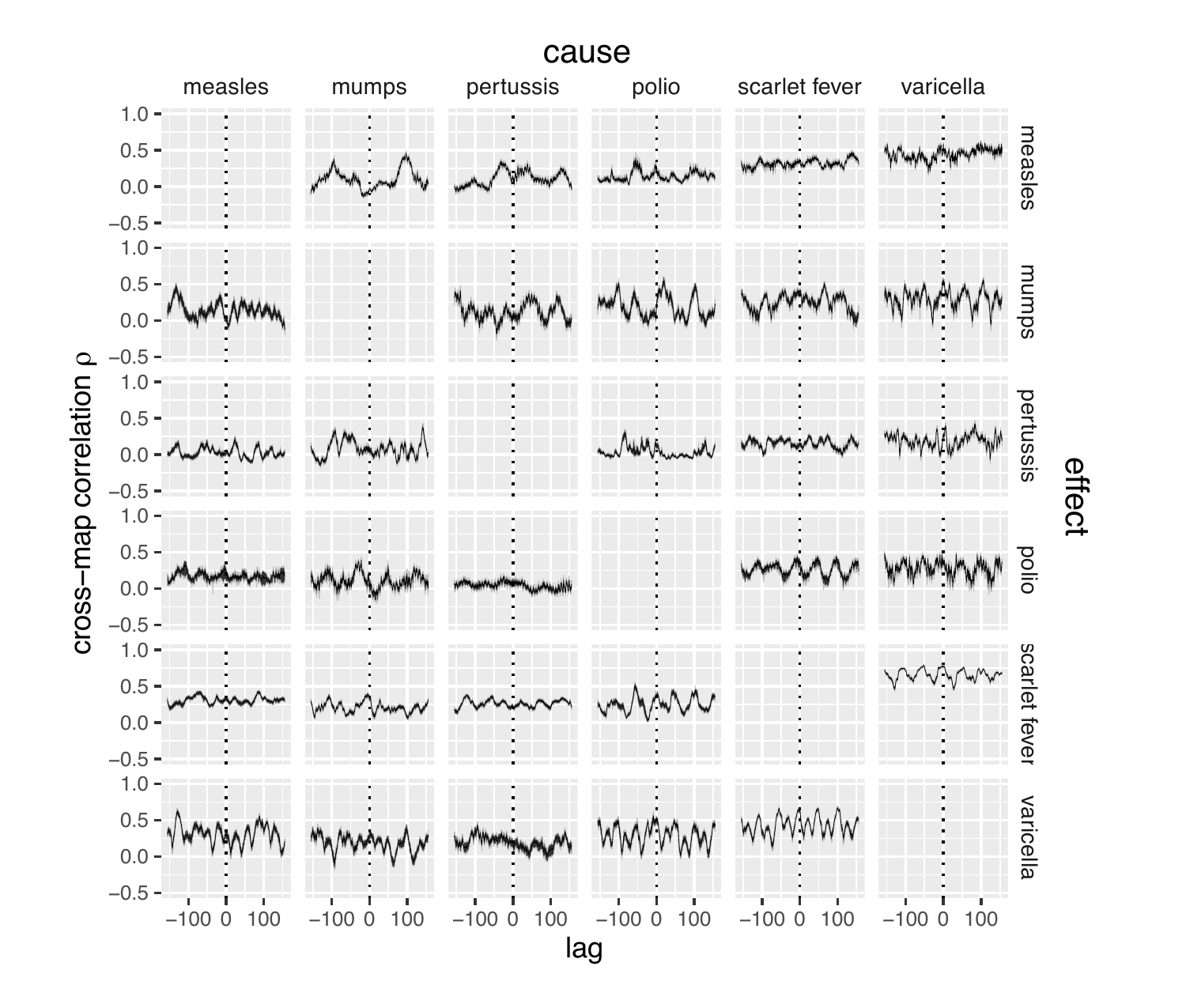}
\end{center}
\caption{\textbf{Cross-map lags for Chicago with default (univariate) embedding}.  \label{fig:cities_corrbylag_chi_self_uniform}}
\end{figure}

Although we specifically chose infectious diseases not subject to major public health interventions in the sampling period, it is possible that the New York data contain noise and transient dynamics.
To the check robustness of the conclusions, we analyzed analogous time series from Chicago from the same period (Fig.~\ref{fig:historical_data_tmp}B).
Completely different interactions appeared (Fig.~\ref{fig:historical_data_tmp}C).
Not correcting for multiple tests, pertussis drove scarlet fever and varicella; accepting marginally significant negative lags ($p=0.055$), polio drove measles.
In these cases, the maximum cross-map correlation $\rho$ was not only positive but also increased at negative lag.
Requiring that $\rho$ only be positive at negative lag, polio also drove pertussis, measles drove mumps and varicella, and mumps drove scarlet fever.
Except in one case, all negative lags occurred at more than one year (Fig.~\ref{fig:cities_corrbylag_chi_self_uniform}).
Thus, no consistent interactions appeared in epidemiological time series of two major, and possibly dynamically coupled, cities.

\begin{figure}
\begin{center}
\includegraphics[width=6in]{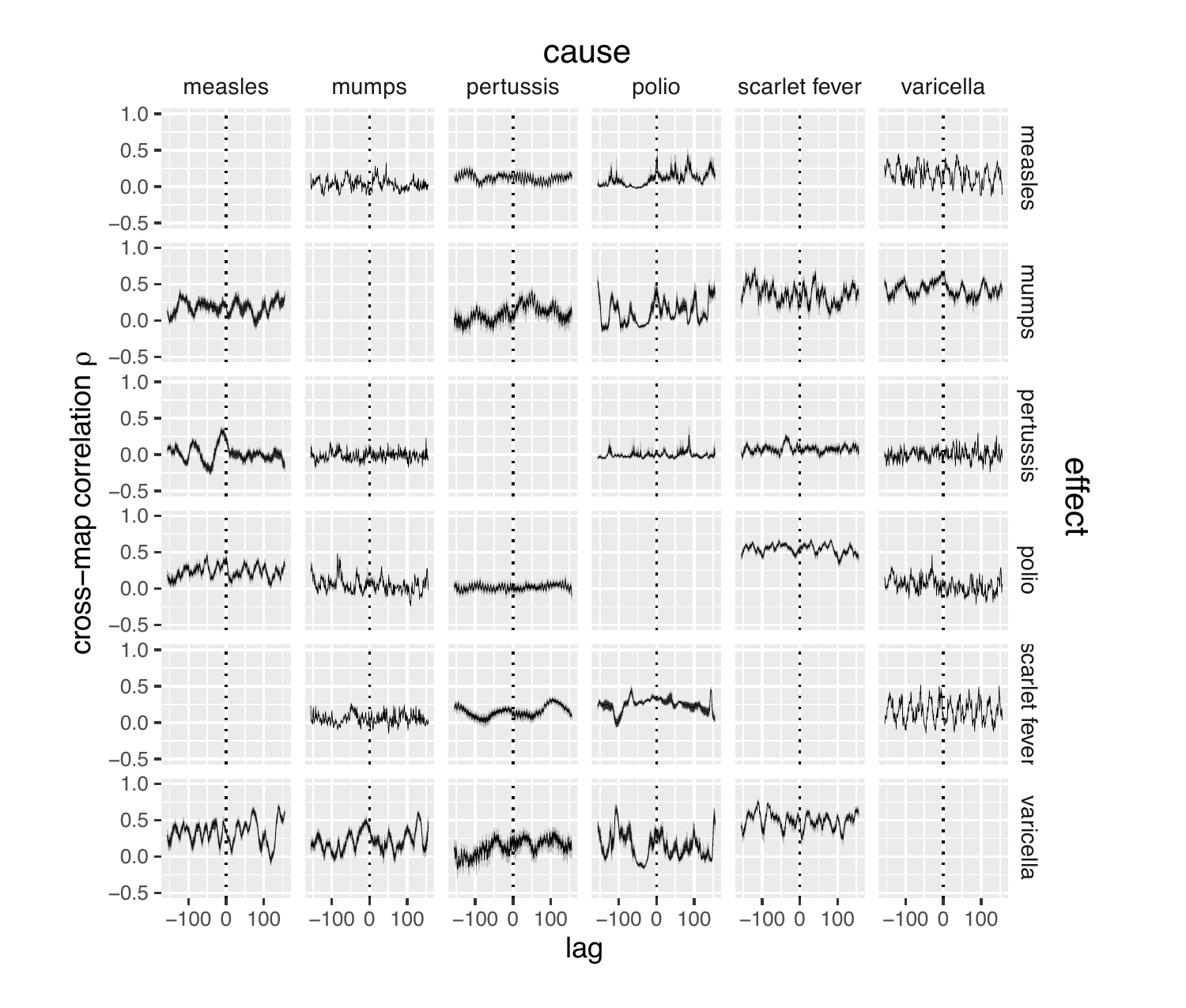}
\end{center}
\caption{\textbf{Cross-map lags for New York with embedding based on random projection}.  \label{fig:cities_corrbylag_nyc_cross_projection}}
\end{figure}

\begin{figure}
\begin{center}
\includegraphics[width=6in]{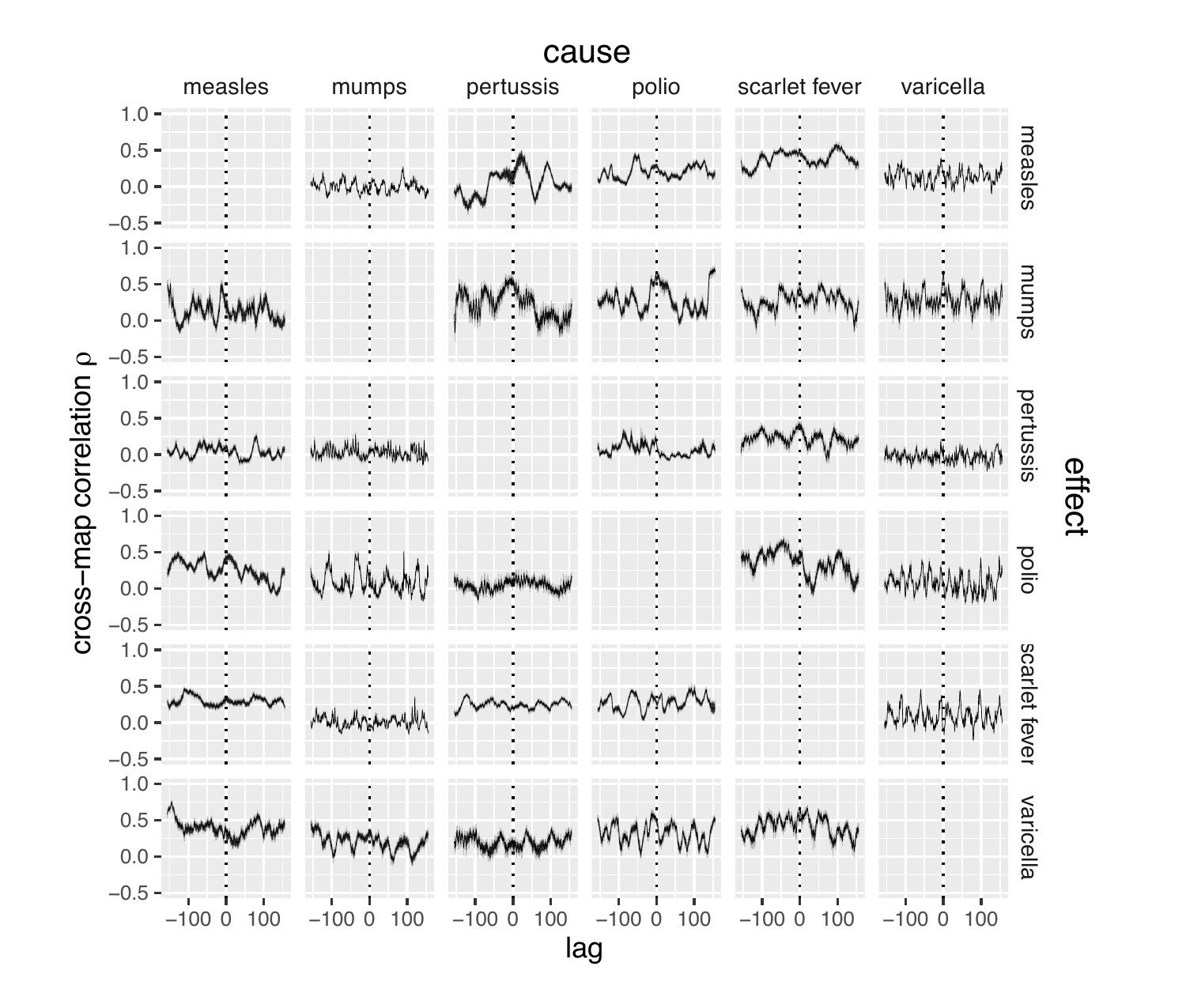}
\end{center}
\caption{\textbf{Cross-map lags for Chicago with embedding based on random projection}.  \label{fig:cities_corrbylag_chi_cross_projection}}
\end{figure}

To investigate the possibility that our method of attractor reconstruction might be unduly sensitive to noise and transient dynamics, we repeated the procedure with a method based on random projections \cite{Tajima2015}.
Once again, no interactions were common to both cities (Fig.~\ref{fig:historical_data_tmp}D).
Furthermore, only one of the original eight interactions from the first reconstruction method reappeared with random projection (two of eight reappeared if disregarding the city), and two interactions changed direction (three if disregarding the city).
Both reconstruction methods selected similar lags (Figs.~\ref{fig:cities_corrbylag_nyc_cross_projection},~\ref{fig:cities_corrbylag_chi_cross_projection}).

\section*{Discussion}

CCM is, in theory, an efficient alternative to mechanistic modeling for causal inference in nonlinear systems.
By evaluating properties of reconstructed dynamics in state space, it sidesteps any need to formulate and fit what are often inaccurate mathematical models.
In current practice, CCM appears an unstable basis for inference in natural systems.
We simulated two interacting strains and found that the original CCM (Criterion 1) can lead to erroneous conclusions whenever strains fluctuated at similar frequencies.
Applying a different criterion for causality that considers the temporal lag at which the cross-map correlation is maximized \cite{Ye2015}, rather than the change in the cross-map correlation with time series length $L$ \cite{Sugihara2012}, avoids this problem.
Inference with Criterion 2 is somewhat robust to process noise, which can improve performance in some cases.
But the method has two problems, even with perfect and abundant observations.
First, it remains susceptible to deviations from its core dynamical assumptions.
``High'' process noise and transient dynamics each diminish performance, leading to false positives and negatives.
Although some observed systems may follow deterministic dynamics that do not themselves change in time, this assumption is often dubious in ecology.
Second, even when the dynamical assumptions are upheld, seemingly equally justifiable methods of attractor reconstruction yield different results.
If the aim is to test hypotheses statistically, these problems raise doubts about the suitability of methods based on state-space reconstruction in ecology.

Oscillations are common in nature, especially in infectious diseases, and suggest that the original criterion (Criterion 1) for causal inference could routinely mislead.
Climatic and seasonal cycles, driven by such factors as school terms, El Ni{\~{n}}o, and absolute humidity, pervade the dynamics of many pathogens and influence the timing of epidemics \cite{Shaman2010, Laneri2010, Finkenstadt2000, Altizer2006, Metcalf2009}.
Infectious diseases can also exhibit fluctuations in the absence of external forcing.
These fluctuations arise from transient damped oscillations or from noise, which induces fluctuations on characteristic time scales and can interact with seasonal drivers to generate complex patterns \cite{Alonso2006, Nguyen2008, Rand1991, Rohani2002}.
Consumer-resource interactions \cite{Boland2009, McKane2005, Turchin2003} and patchy populations \cite{Nisbet1978, Durrett1994} demonstrate similar behavior.
In systems with synchronized dynamics, the only demonstrated reliable criterion for causal inference is a negative cross-map lag \cite{Ye2015}.

Assuming the stronger criterion for causality \cite{Ye2015}, under what conditions might we consider this method ``safe''?
We have shown that departures from a fixed attractor are a problem.
These departures constitute different forms of transient dynamics.
From a modeling perspective, we could describe them as arising from initial conditions, process noise, or a change in the underlying attractor due to a secular change in a parameter.
In our system, a $\geq5\%$ standard deviation in the transmission rate generated appreciable false positives.
Is this high or low?
Although the amount of process noise in a model can be estimated by the variance of the dynamics not explained by the deterministic skeleton, if the true skeleton is unknown, estimates are sensitive to the approximating statistical functions \cite{Ellner1995}.
More importantly, the existence of transient dynamics in a time series indicates insufficient observations.
There is furthermore no guarantee any natural system will reach an attractor before going extinct or that the system's dynamics themselves do not evolve \cite{Turchin2003}.

\begin{figure}
\begin{center}
\includegraphics[width=6in]{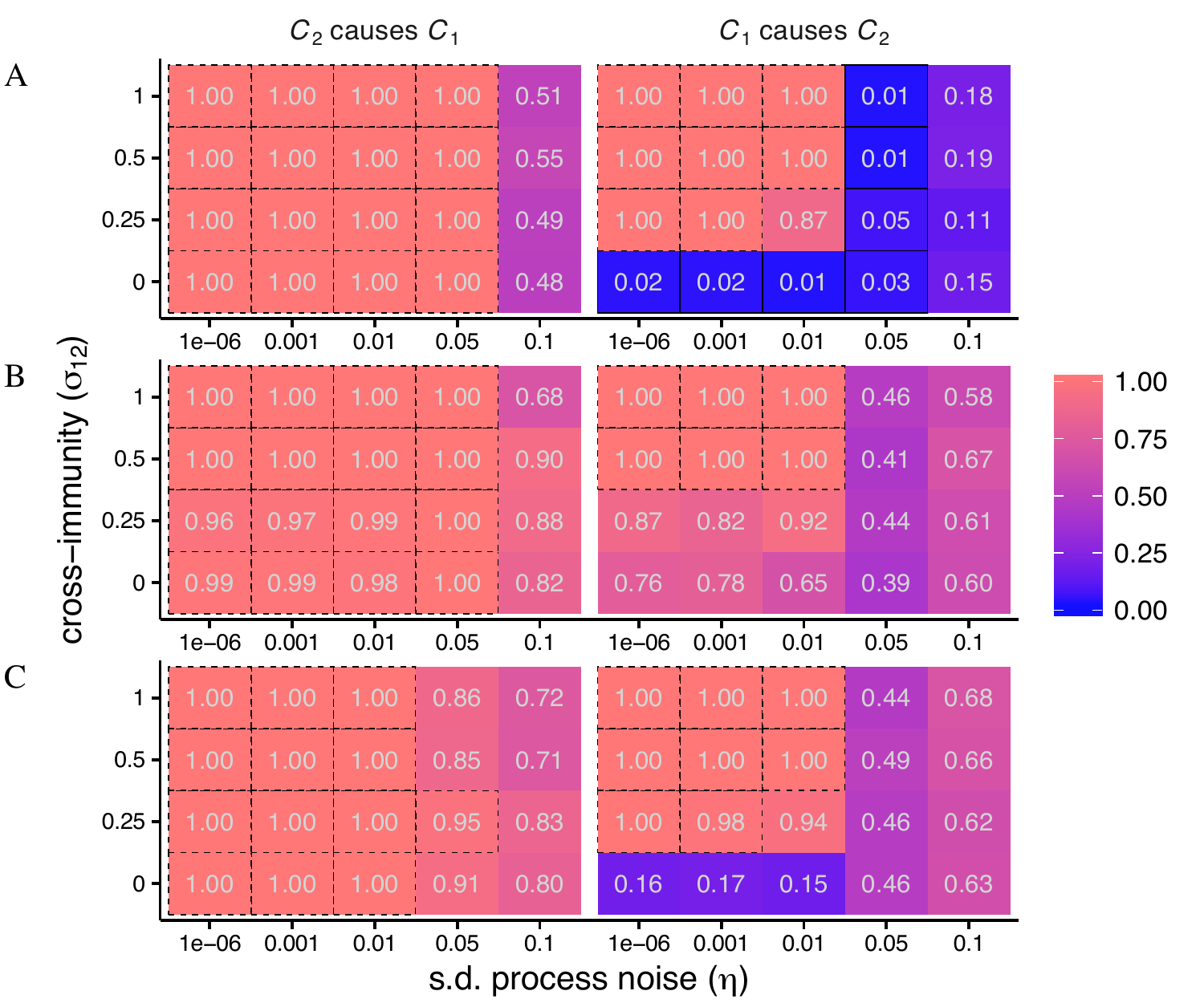}
\end{center}
\caption{\textbf{Interactions detected as a function of process noise and the strength of interaction ($C_2 \rightarrow C_1$) for different delay-embedding methods}. Heat maps show the fraction of 100 replicates significant for each inferred interaction for different parameter combinations. A significant increase in cross-map correlation $\rho$ with library length $L$ indicated a causal interaction. Each analysis is based on 100 years of monthly data. Delay-embeddings were chosen by (A) nonuniform embedding, (B) random projection, or (C) maximizing the cross-map correlation $\rho$.  \label{fig:detect_diffembed}}
\end{figure}

If an ecologist were confident that observed dynamics reflected dynamics near a fixed, deterministic attractor (e.g., in a simple, closed system), uncertainties in the methodology of attractor reconstruction still suggest caution.
We tested four different methods of selecting the lag-embedding.
Even near an attractor, they gave different results (Fig.~\ref{fig:detect_diffembed}).
Decades of research on methods of attractor reconstruction show the continued difficulty of justifying a particular approach \cite{Casdagli1991, Uzal2011, Nichkawde2013, Tajima2015, Sugihara1990}.
Reconstructions from unknown systems thus currently run the risk of being ad hoc and compromising causal inference.
The statistics for evaluating cross-map correlations also deserve attention.
We bootstrapped and attempted to validate approaches empirically with simulated data, but the methods are not rigorously grounded in a probabilistic framework such as those common to mechanistic modeling~\cite{HilbornMangel}.
Extending the approach to explicitly link nonlinear dynamics with process and observation noise in a probabilistic framework has the potential to put the method on a sounder footing.

Of the many factors that might explain the contrasting results for childhood infections in two cities, biological explanations thus seem the least likely.
Although there is evidence that measles increases suceptibility to other pathogens \cite{Mina2015}, and that measles and pertussis compete for susceptible hosts \cite{Rohani2003}, the CCM analyses did not consistently support either hypothesis.
It is difficult to imagine a parsimonious mechanism by which the inferred interactions might be plausible.
Different rates or modes of transmission for each disease in each city might lead to varying patterns of infection in different subpopulations, which would affect interactions.
We know of no support for this hypothesis.
In contrast, we cannot rule out transient dynamics, which could arise from changes in birth rates, mobility, and behavior during this period \cite{Earn2000}.
Process noise, implying the omission of important state variables and poor resolution of the underlying deterministic attractor, could also affect performance.
Errors in attractor reconstruction are another possibility.
Except for pertussis, different delay-embeddings were selected for each pathogen in each city, and an alternative method of attractor reconstruction yielded even more divergent results.
Finally, we cannot account for the effects of short time series and measurement error.
We conclude that the inferred interactions are untrustworthy.

Detecting causality remains challenging in the face of real data from a complex world.
With limited data and complex dynamics, mechanistic models are always misspecified to some extent, and the use of other lines of evidence to motivate the choice of model structure is necessary for good inference \cite{BurnhamAnderson, He2009, Yodzis1988, Wood1999, Grad2012}.
But even an accurate mechanistic model that reproduces observed patterns well cannot prove causality.
Controlled manipulative experiments, which are notoriously hard to conduct in large complex systems, are necessary.
Global systems can never sustain this high standard.
Randomization and replication are often possible on lower scales, but inference is complicated by the fact that replicates may not be truly independent \cite{Simberloff1969, Hurlbert1984, Tilman1989}.
With diseases like the ones we invesigate here, manipulations (e.g., vaccination) are furthermore seldom feasible.
This has led epidemiologists and disease ecologists to resort to a mishmash of heuristics, frequently based on observational data, for causal inference \cite{Plowright2008}.
Prediction, in contrast, is epistemologically straightforward and useful without knowledge of the true underlying structure of a system.
It does not require deciding a priori what the best method is (model-based, model-free, or hybrid): the proof is in the prediction.
Predictive and mechanistic models may converge if the predictive factors are chosen to mimic the hypothesized state variables over time.

Beyond its statistical practicalities, the prospect of applying state-space reconstruction to causal inference touches on unsettled questions in ecology.
Are systems approximately deterministic and settled on static attractors, and how can we tell?
Although CCM does not require that dynamics follow an identifiable model, it does require sufficient coverage of a fixed state-space \cite{Hastings2004}.
We propose that this position is justifiable only in systems that are already well-understood (e.g., closed, non-evolving microcosms at steady state), but in these cases, causality is typically known.

\section*{Methods}

\subsection*{Dynamical model}
\begin{figure}
\begin{center}
\includegraphics[width=5in]{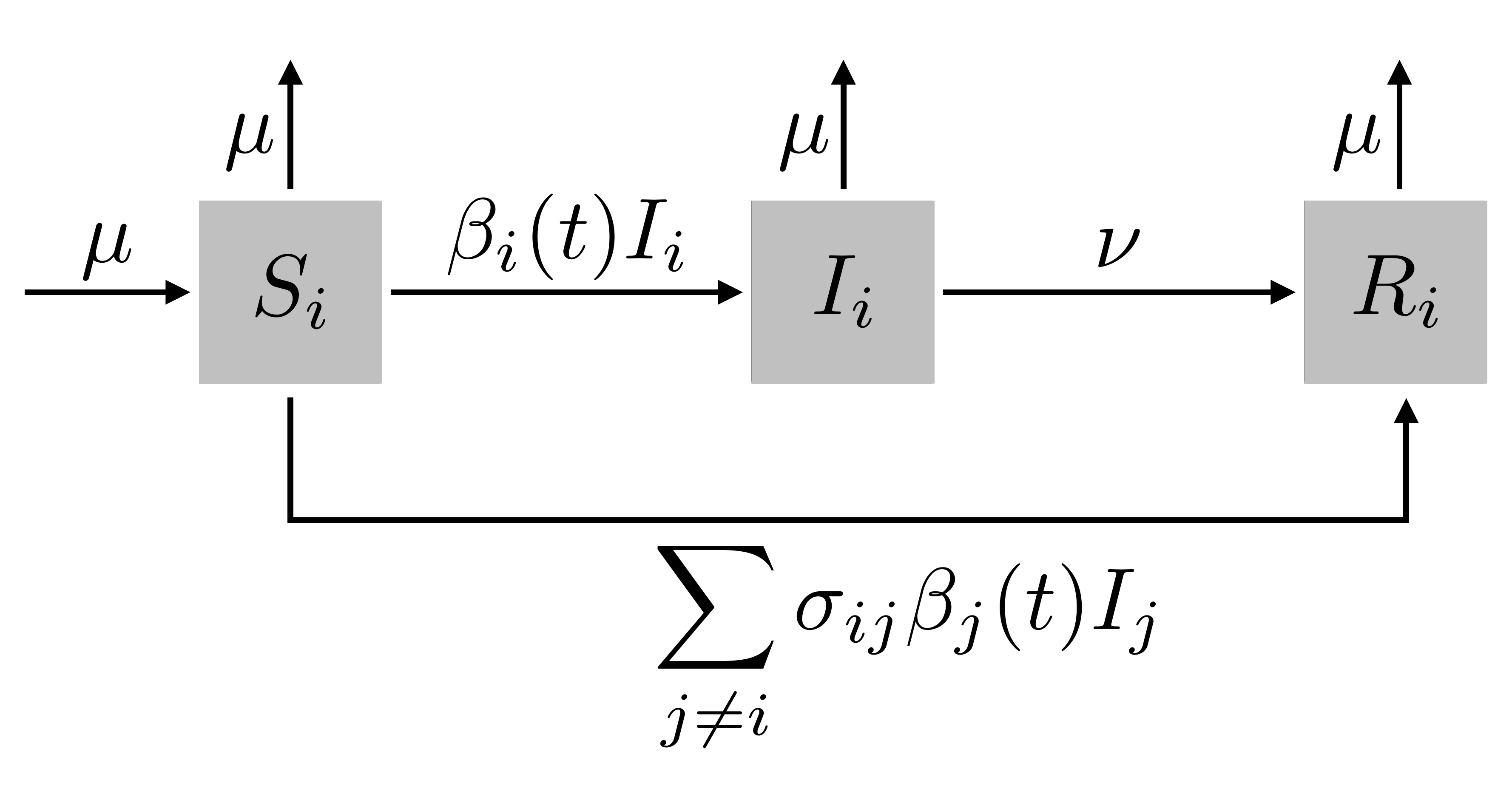}
\end{center}
\caption{\textbf{Compartmental representation of strain-competition model}.
Hosts are susceptible (S), infected/infective (I), or recovered (R) with respect to each strain.
Hosts move from S to I based on a seasonally varying transmission rate, and from I to R at a constant recovery rate. Competition takes place through cross-immunity, which is implemented by having hosts skip the infected state for one strain with some probability if they are already infected with another strain.
\label{fig:compartmental}}
\end{figure}

We modeled the dynamics of two pathogen strains under variable amounts of competition and process noise (Fig.~\ref{fig:compartmental}).
The state variables in the system are the hosts' statuses with respect to each strain \cite{Gog2002}.
Hosts can be susceptible ($S_i$), infected ($I_i$), or recovered and immune ($R_i$) to each strain $i$.
The deterministic model has the form:

\begin{align}
\frac{dS_i}{dt} &=
\mu
- S_i\sum\limits_{j}
\sigma_{ij}
\beta_j(t)
I_j
- \mu S_i \\
\frac{dI_i}{dt} &=
\beta_i(t) S_i I_i
- (\nu_i + \mu) I_i \\
\frac{dR_i}{dt} &=
\nu_i I_i
+ S_i\sum\limits_{j \neq i} \sigma_{ij} \beta_j(t)  I_j
- \mu R_i \\
\beta_i(t) &= \beta_i
\left(
1 + \varepsilon \sin \left[
\frac{2\pi}{\psi} \left( t - \psi \right)
\right]
\right) \\
S_i + I_i + R_i &= 1
\end{align}

Hosts enter the susceptible class for strain $i$ through the birth (and death) rate $\mu$.
They leave through infection with strain $i$ ($S_i \to I_i$), infection with strain $j$ that elicits cross-immunity to $i$ ($S_i \to R_i$), or death.
The per capita transmission rate, $\beta_i(t)$, depends on a mean strain-specific rate, $\beta_i$, and a forcing function that is shared by all strains.
This function has a sinusoidal form and represents a shared common driver, such as seasonal changes in susceptibility or transmission from school-term forcing.
The forcing function is defined by a shared period $\psi$ and amplitude $\epsilon$.
Infected hosts recover at rate $\nu_i$ ($I_i \to R_i$).
The immune host class grows through these recoveries and also from the fraction of susceptible hosts, $S_i$, contacting infected hosts, $I_j$, who develop cross-immunity, $\sigma_{ij}$ ($0 \leq \sigma_{ij} \leq 1$).
Immunity of this form has been described as ``polarizing'' because $\sigma_{ij}$ of hosts $S_i$ contacting infecteds $I_j$ become completely immune (non-susceptible) to strain $i$, while $1-\sigma_{ij}$ remain completely susceptible.
This cross-immunity is a form of competition that determines the directions of interaction between strains: when $\sigma_{ij}>0$, strain $j$ drives strain $i$.
We assume $\sigma_{ii}=1$: hosts acquire perfect immunity to a strain from which they have recovered.

Process noise on the per capita transmission rate produces stochastic differential equations in Ito form:
\begin{align}
dS_i &=
[\mu - \mu S_i] \, dt
- S_i\sum\limits_{j} \sigma_{ij} \beta_j(t) I_j [dt +\eta \, dW_{t,j}]
\\
dI_i &=
\beta_i(t) S_i I_i [dt +  \eta \, dW_{t,i}]
- [\nu_i + \mu] I_i \,  dt \\
dR_i &=
[\nu_i I_i - \mu R_i] \, dt
+ S_i\sum\limits_{j \neq i} \sigma_{ij} \beta_j(t) I_j [dt +  \eta \, dW_{t,j}]
\end{align}
where the $W_i$ are independent Wiener processes, one for each pathogen $i$, and $\eta$ represents the standard deviation of the noise as a fraction of the deterministic transmission rate.

The observations consist of the number of new cases or incidence over some interval.
Cumulative cases $c_i$ at time $t$ were obtained by summing the $S_i \to I_i$ transitions from the start of the simulation through time $t$.
The incidence over times $t-\Delta t_\text{obs}$ to $t$, written as $C(t)$ for convenience, is given by the difference in cumulative cases:
\begin{align}
C_i(t) &= c_i(t_2) - c_i(t_1) \\
dc_i &= \beta_i(t) S_i I_i [dt + \eta \, dW_{t,i}]
\end{align}

\begin{table}
\caption{Default parameter values.}
\begin{center}
\begin{tabular}{cccc}
{\bf Symbol} &{\bf Description} & {\bf Default value} \\
\hline
$\beta_1, \beta_2$ & transmission rates & 0.3, 0.25 $\text{d}^{-1}$ \\
$\sigma_{12}$ & immunity to strain 1 from infection with 2 & see text\\
$\sigma_{21}$ & immunity to strain 2 from infection with 1 & 0\\
$\sigma_{ii}$ & homologous immunity for strain $i$ & 1\\
$\mu$ & birth and death rate & 1/30 $\text{y}^{-1}$ \\
$\nu$ & recovery rate & 0.2 $\text{d}^{-1}$ \\
$\epsilon$ & amplitude of seasonal forcing & 0.1 \\
$\psi$ & period of seasonal forcing & 360 d\\
$\eta$ & standard deviation of process noise & see text \\
$\text{S(0)}$ & initial fraction susceptible & see text \\
$\text{I(0)}$ & initial fraction infected & see text\\
$\Delta t_\text{obs}$ & incidence and sampling interval & 30 days\\
\end {tabular}
\end{center}

\label{table:defaultParameters}
\end{table}

\subsection*{Simulation}

The equations were solved numerically using the Euler-Maruyama method with a fixed step size.
The step size was chosen to be less than the smallest within-run harmonic mean step size across deterministic, adaptive-step size pilot runs performed across the range of parameter space being studied.
When numerical errors arose during transients, the step size was reduced further until the numerical issues disappeared.

Except where noted, the model was simulated with random initial conditions, and 1000 years of monthly observations were obtained from stochastic fluctuations around the deterministic attractor.
The use of random initial conditions minimizes arbitrary bias in the simulated dynamics.
From visual inspection of dynamics, the transient phase lasted much less than 1000 years. Time series were obtained from years 2000-3000.

\subsection*{Cross-mapping}

Convergent cross-mapping (CCM) is a method for inferring causality in deterministic systems via delay embedding~\cite{Sugihara2012}.
Takens' theorem holds that, for an $E$-dimensional system, the attractor for the state space represented by delay vectors in a single variable $X$, $\mathbf{x}(t) = \{X(t), X(t - \tau_1), X(t - \tau_2), \ldots , X(t - \tau_{E-1}) \}$, is topologically equivalent to the $E$-dimensional attractor for variables $X_1, ..., X_E$.
In the limit of infinite data, the full $E$-dimensional attractor can be reconstructed perfectly from a one-dimensional time series.
Therefore, because $\bx(t)$ contains complete information about the system's dynamics, if $Y$ is part of the same system and thus causally drives $X$, observations of $\bx(t) \rightarrow Y(t - \ell)$, for a fixed lag $\ell$, can be used to reconstruct unobserved values of $Y(t)$ from new observations of $\bx(t)$~(Fig.~\ref{fig:conceptual}).

To evaluate whether $Y$ drives $X$, we construct ``libraries'' of observations of $\bx(t) \rightarrow y(t - \ell)$.
For a particular library, we treat each value of $Y(t)$ as unobserved, and reconstruct its value $\hY(t)$ by identifying the $E + 1$ nearest neighbors to $\bx(t)$ in the library, $\bx(t_i)$, for $t_1, \ldots, t_{E+1}$, and calculating $\hY = \sum_{i = 1}^{E + 1} w_i Y(t_i)$.
In order to avoid predictability due to system autocorrelation rather than dynamical coupling, neighbors are restricted to be separated in time by at least three times the delay at which the autocorrelation drops below $1/e$.
Weights are calculated from the Euclidean distances $d_i$ between $\bx(t)$ and $\bx(t_i)$, with $w_i$ proportional to $\exp \left( -\frac{d_i}{d_0} \right)$, where $d_0$ is the distance to the nearest neighbor \cite{Sugihara2012}.

The cross-map correlation $\rho$ measures how well values of $Y$ can be reconstructed from values of $X$, and is defined as the Pearson correlation coefficient between reconstructed values $\hY(t)$ and actual values $Y(t)$ across the entire time series \cite{Ye2015}.
Given library size $L$ and lag $\ell$, we generate a distribution of cross-map correlations $\rho$ by bootstrap-sampling libraries mapping delay vectors $\bx(t)$ to values $Y(t - \ell)$ and then computing the cross-map correlation for each sampled library.
We use the bootstrap distribution of cross-map correlation as the basis for statistical criteria for causality.

\subsection*{Criteria for causality}

We infer causality using two primary criteria involving the cross-map correlation $\rho$ \cite{Sugihara2012, Ye2015}: (1) whether $\rho$ increases with $L$ for a fixed lag $\ell$, and (2) whether $\rho$ is positive and maximized at a negative temporal lag $\ell$.
We also consider a weaker alternative to the first criterion, which is simply whether $\rho$ is positive.

\paragraph{Criterion 1}
If $Y$ drives $X$, then increasing the library size $L$ should improve predictions of $\bx(t)$ as measured by $\rho$~\cite{Sugihara2012} for fixed lag $\ell = 0$.
The first criterion tests for this increase in $\rho$ with $L$.
We calculate $\rho$ at $L_{\min} = E + 2$, the smallest library that will contain $E + 1$ nearest neighbors for delay vectors $\bx(t)$, and at $L_{\max}$, the total number of delay vectors $\bx(t)$ in the time series.
An increase in $\rho$ is indicated by a lack of overlap between the distributions at $L_{\min} = E + 2$, the smallest library that will have $E + 1$ neighbors for most points, and $L_{\max}$, the largest possible library given the time-series length and delay embedding parameters $E$ and $\tau$.

\paragraph{Criterion 2}
If $Y$ strongly drives $X$, cross-map correlation at $\ell = 0$ may yield a false positive when testing for $X$ driving $Y$, but because information is transferred forwards in time from $Y$ to $X$, the cross-map correlation should be maximized at a negative lag $\ell$~\cite{Ye2015}.
The second criterion simply requires that, to infer that $Y$ drives $X$, the cross-map correlation $\rho$ be maximized at a negative cross-map lag $\ell$ and be positive.
In other words, not only must $X$ contain information about $Y$, but this information must be greatest for past states of $Y$, reflecting the correct temporal direction for causality.

\subsection*{Statistical tests for causality criteria}

The theory underlying CCM assumes completely deterministic interactions and infinite data.
If $Y$ drives $X$ in the absence of noise, the correlation $\rho$ between the reconstructed and observed states of $Y$ should converge to one with infinite samples of $X$.
In practice, if $X$ and $Y$ share a complex (e.g., chaotic) attractor, time series of $X$ may not be long enough to see convergence~\cite{Sugihara2012}.

The presence of observation and/or process noise violates the deterministic assumptions and prevents $\rho$ from ever reaching 1.
Nonetheless, a detectable increase in the correlation $\rho$ with the library length $L$ (for Criterion 1), or a maximum and positive correlation at negative lag (for Criterion 2), may suffice to demonstrate that $X$ drives $Y$ in natural systems.
It is important to note that we have no formal theoretical justification for such statistical heuristics.

Our statistics are based on the distributions obtained from bootstrapping.
For Criterion 1, which tests for an increase in $\rho(L)$, we perform a nonparametric test of whether $\rho(L_{\max})$, obtained at the largest library length is greater than $\rho(L_{\min})$, obtained at the smallest libary length.
The p-value for this test is calculated as the probability that $\rho(L_{\max})$ is not greater than $\rho(L_{\min})$, and calculate the p-value directly from the sampled distributions (the fraction of bootstraps in which $\rho(L_{\max}) < \rho(L_{\min})$).
We also consider a weaker alternative, testing simply whether $\rho$ is significantly positive.

For Criterion 2, which tests whether the best cross-map lag is negative and thus indicates the correct causal direction in time, we perform a similar nonparametric test.
We identify the negative cross-map lag $\ell^{(-)}$ with the highest median correlation, $\rho(\ell^{(-)})$ as well as the nonnegative cross-map lag $\ell^{(0+)}$ with the highest median correlation.
The p-value for this test is calculated as the probability that $\rho(\ell^{(-)})$ is not greater than $\rho(\ell^{(0+)})$.

We use a significance threshold of $p<0.05$ for all tests.

\subsection*{Choice of delay and embedding dimension}
The theory underlying attractor reconstruction works with any $E$-dimensional projection of a one-dimensional time series, which can be generated in many ways from lags of the time series.
In simulated, deterministic models, $E$ can be known perfectly, but the best projection may be system-dependent.
In systems with process noise, unknown dynamics, and/or finite observations, there is no clearly superior method to select the appropriate projection \cite{Casdagli1991,Nichkawde2013,Uzal2011,Pecora2007, Cao1997, Small2004}.

We accommodated this uncertainty by using four different methods.
Two methods infer the best delay-embedding for each interaction by maximizing the ability of one variable, the driven variable, to predict itself (akin to nonlinear forecasting \cite{Sugihara1990, Sugihara1994}).
The third method instead uses the delay-embedding that maximizes the cross-mapping correlation $\rho$ for each interaction.
Three of the four methods use uniform embeddings, identifying $E$ and a fixed delay $\tau$, and the other uses a nonuniform embedding, identifying a series of specific delays $\tau_1$, $\tau_2$, etc., whose length determines $E$.

\begin{enumerate}
\item \textit{Univariate prediction method}: By default, for each causal interaction ($C_i \rightarrow C_j$), $E$ and $\tau$ are chosen to maximize the one-step-ahead univariate prediction $\rho$ at $L_{\max}$ for the driven variable ($C_j$) based on its own time series.
\item \textit{Maximum cross-correlation method}: As an alternative, $E$ and $\tau$ are chosen to maximize the mean cross-map correlation $\rho$ at $L_\text{max}$ for each causal interaction being tested, for each time series.
\item \textit{Random projection method}: A recently proposed method based on random projection of delay coordinates sidesteps the problem of choosing optimal delays \cite{Tajima2015}. Instead, for a given $E$, all delays up to a maximum delay $\tau_{\max}$ are projected onto an $E$-dimensional vector via multiplication by a random projection matrix. $E$ is chosen to maximize the cross-map correlation $\rho$.
\item \textit{Nonuniform method}: For each driven variable $C_j$, starting with $\tau_0 = 0$, additional delays $\tau_1, \tau_2, \ldots$ are chosen iteratively to maximize the directional derivative to nearest neighbors when the new delay is added \cite{Nichkawde2013}. The delays are bounded by the optimal uniform embedding based on a cost function that penalizes irrelevant information \cite{Uzal2011}. This method can be seen as a nonuniform extension of the method of false nearest neighbors \cite{Kennel1992}.
\end{enumerate}

\subsection*{Code}
Code implementing the state-space reconstruction methods is publicly available at \url{https://github.com/cobeylab/pyembedding}.
The complete code for the analysis and figures is publicly available at \url{https://github.com/cobeylab/causality_manuscript}; individual analyses include references to the Git commit version identifier in the `pyembedding' repository.
The simulated time series on which the analyses were performed are available from the authors on request.

\subsection*{Data on childhood infections}
Time series were obtained from L2-level data maintained by Project Tycho \cite{vanPanhuis2013}.
All available cases of measles, mumps, pertussis, polio, scarlet fever, and varicella were obtained from the first week of 1906 through the last week of 1953 for New York City and Chicago.
Pertussis data were terminated in the 26th week of 1948 to limit the influence of the recently introduced pertussis vaccine.
Incidence was calculated by dividing weekly cases by a spline fit to each city's population size, as reported by the U.S. Census.

\section*{Acknowledgements}

We thank Mercedes Pascual, Lauren Childs, and Greg Dwyer for helpful comments.

\subsection*{Funding}

This work was supported in part by the University of Chicago Big Ideas Generator, and was completed in part with resources provided by the University of Chicago Research Computing Center.

\bibliography{ccm_ms}

\beginsupplement

\section*{Supplement}

\end{document}